\documentclass[a4paper,11pt]{article}

\usepackage{jheppub} 

\usepackage{xcolor} 

\title{\boldmath   Momentum analyticity of the holographic electric polarizability in 2+1 dimensions }

\author[a,c]{Lei Yin} 
\author[b,c]{Hai-cang Ren} 
\author[a]{Ting-Kuo Lee} 
\author[c]{Defu Hou}

\affiliation[a]{Institute of Physics, Academica Sinica ,\\Taipei   11529, R.O.C.} 
\affiliation[b]{Physics Department, The Rockefeller University,  \\ 1230 York Avenue, New York, 10021-6399, U.S.A.} %
\affiliation[c]{Institute of Particle Physics and Key Laboratory of Quark and Lepton Physics (MOS) ,  \\ Central China Normal University,  Wuhan,  430079, P.R.C.} %

\emailAdd{leiyinbox@gmail.com} 
\emailAdd{ren@mail.rockefeller.edu} 
\emailAdd{tklee@phys.sinica.edu.tw} 
\emailAdd{houdf@mail.ccnu.edu.cn}

\abstract{ 
The static electric polarization of a holographic field theory dual to the Einstein-Maxwell theory in the background of $AdS_4$ with a Reissner-Nordstr\"{o}m (AdS-RN)  black hole is investigated.  
We prove that the holographic polarization is a meromorphic functions in complex momentum plane and locate analytically the asymptotic distribution of the poles along two straight lines parallel to 
the imaginary axis for a large momentum magnitude. The results are compared with the numerical result on Friedel-like poles of the same holographic model reported in the literature and with 
the momentum singularities of the one-loop polarization in weak-coupling spinor QED$_3$ and scalar QED$_3$ with the similarities and differences discussed.
}


\begin{document} 
\maketitle
\flushbottom 

\section{Introduction} 
\label{sec:introduction}

The gauge/gravity duality \cite{Witten1998b,Maldacena1998,Aharony1999,Klebanov1999,witten2001multi} has evolved into an important tool to explore the 
strongly interacting systems in high energy physics and condensed matter physics. By converting the path integral of a strongly interacting 
quantum system into the classical theory of a weakly coupled gravity-matter system with one more space dimension, the duality opens a new avenue towards 
a qualitative or even quantitative understanding of the non-perturbative effects \cite{Maldacena1998b,Policastro2001}, especially when the problem cannot 
be tackled by numerical techniques. 
The holographic superconductor provides such an example, where the quantum effective action of a strongly coupled superconductor in 2+1 dimensions corresponds to a 
3+1 dimensional classical action of an Abelian-Higgs theory coupled to the gravity with a AdS-Reissner-Nordstr\"{o}m (AdS-RN) black hole \cite{Gubser2008}, 
or with a AdS- Schwarzschild black hole in the probe limit \cite{Hartnoll2008,Yin2015}. 
The ratio between the absorption threshold of the AC conductivity (the onset frequency of its real part) and the critical temperature
extracted from the holographic superconductor turns out 
to be close to the observed values from cuprates superconductors\cite{Horowitz2008,Hartnoll2008a}.

Unlike the Type IIB superstring theory in $AdS_5\times S^5$, which corresponds to $\mathcal{N}=4$ super Yang-Mills in 3+1 dimensions,
all holographic models used in condensed matter physics follow a bottom-up approach without the knowledge of the explicit Lagrangian underlying the quantum effective 
action implied by the gauge/gravity duality, as is reflected, for instance, in the lack of the evidences such as the Andreev reflection that link the order parameter 
in the holographic superconductivity to the Cooper pairing of the fermionic degrees of freedom if any. On top of all its applications, the duality itself remains 
a conjecture. Therefore in addition to conquer more strongly correlated systems with 
more sophisticated holographic models, it is equally important to collect more fundamental properties shared by existing holographic models and  
ordinary field theories, and thereby to accumulate more evidences supporting the conjectured duality.
The work reported below serves the latter purpose and the holographic model to be addressed is the Einstein-Maxwell theory in 3+1 AdS-Reissner-Nordstr\"{o}m (AdS-RN) black 
hole background, which describes the normal phase of  the 2+1 dimensional holographic superconductor\cite{Yin2016}. The analyticity of the electric component of the static 
polarization tensor $\Pi_{\mu\nu}(q)$ with respect to the spatial momentum $q$ will be 
examined and the asymptotic distribution of the complex singularities underlying the Friedel oscillations of the dressed Coulomb potential will be obtained and compared 
with the  both 2+1 spinor QED and scalar QED in weak coupling. In this regard, our result is of only theoretical values. Its phenomenological implications remain to be 
unveiled. 

This research is the continuation of the previous one \cite{Yin2016}, where the momentum analyticity of the magnetic component of the polarization tensor (the spatially transverse
component satisfying $q^i\Pi_{ij}(q)=0$) of the same 
holographic model was investigated. There we proved that the static magnetic polarization function is a meromorphic function of the complex spatial momentum. Using the 
WKB approximation, we were able to show that the poles for large momentum magnitude are distributed asymptotically along two lines parallel to the imaginary axis of the 
complex momentum plane. The parallel analyses will be extended to the more interesting electric component of the polarization tensor, $\Pi_{00}(q)$ below. 
The Friedel oscillation in this case is speculated to support the Cooper pairing in some strongly correlated system \cite{Kohn1965} .  
Technically, the polarization tensor is extracted from the solutions of the linearized Einstein-Maxwell equations in the RN blackhole background. The metric tensor and gauge 
potential fluctuations can be divided to two groups according to the parity under the reflection with respect to the line perpendicular to the spatial momentum on the AdS 
boundary. The fluctuations underlying the magnetic component, treated in \cite{Yin2016}, belong to the odd parity group and they are the solutions of the  two coupled Einstein-Maxwell equations in 
the static limit. The fluctuations underlying the electric component belong to the even parity group, which involve four coupled Einstein-Maxwell in the static limit, and
their  analytic treatment becomes more challenging. Nevertheless, after some twist and turns, we are able to reach similar conclusions as the magnetic component. The electric  
component $\Pi_{00}(q)$ is a meromorphic function of the spatial momentum $q$ with poles distributed asymptotically along two lines parallel to the imaginary 
axis of the complex $q$-plane for large $|q|$. The asymptotic locations of these poles match well with those extracted from the numerical solution of the Einstein-Maxwell equations 
for large ${\rm Im}q$ \cite{Blake2015}.

The presence of the complex momentum singularities of the Green's function appears a common property of a field theory, either strongly coupled or weakly coupled, with a nonzero 
chemical potential as demonstrated in this work and the previous ones \cite{Blake2015, Yin2016, Edalati2010}. It is not, however, a sufficient evidence of fermionic degrees of freedom present in 
the boundary field theory of the gauge/gravity dual. The one-loop polarization tensor of  scalar QED3 with a nonzero chemical potential displays those complex momentum singularities as well, 
similar to spinor QED.  

This paper is organized as follows. The momentum analyticity of the polarization tensor of both spinor and scalar QED3 to one-loop order will be discussed in the next
section with the result as a benchmark for comparison with its holographic counterpart. In section 3, the electric component of the holographic polarization tensor will be extracted from the 
even parity solutions of the linearized Einstein-Maxwell equations in AdS background, and the proof of its meromorphism  in a complex spatial momentum will be given. 
The WKB approximation will be employed in the section 4 to find out the asymptotic distribution of the poles on the complex momentum plane. In Section 5, the analytic result obtained 
in this work will be compared with the numerical result of \cite{Blake2015} and conclude the paper.

\section{ One-loop Photon self-energy in $2+1$-dimensional spacetime }
\label{sec:one-loop-correction}

To study the finite-temperature quantum electrodynamics, we start with the functional integral formalism:
\begin{align}
  \mathcal{Z} = \int \mathcal{D}(\phi_\alpha)\, \mathrm{e}^{- S_E}  = \int \mathcal{D}(\phi_\alpha) \exp \Bigg\{ -  \int_0^\beta \ \mathrm{d}\tau \int \ \mathrm{d}^3 \vec{r} \, \mathcal{L}[\phi_\alpha]\Bigg\}   \label{eq:1}
\end{align}
where $S_E$ is the Euclidean action and $\phi_\alpha, \; \alpha = 1,2, \cdots$ represents all fields under consideration. 

\subsection{ Fermion case: Spinor QED }
\label{sec:fermion-case:-spinor}

We consider first the spinor QED, and the Lagrangian density in (\ref{eq:1}) is chosen as 
\begin{align}
    \mathcal{L}[\psi,\bar\psi]=-\bar\psi\gamma_\lambda\left(\frac{\partial}{\partial x_\lambda}-ieA_\lambda\right)\psi+\mu\bar\psi\sigma_3\psi
    \label{lagrange}
\end{align}
where $\psi$ and $A_\lambda$ are the massless fermion field and $U(1)$ gauge potential in $2+1$ dimensional space-time, respectively,  $e$ is the electric charge and $\mu$ is the chemical potential.  The representation of $\gamma$ matrices in $2+1$ dimensional space-time reads
\begin{align}
  \gamma_0 = \sigma_3 , \quad \gamma_1 = - \mathrm{i} \sigma_1, \quad \gamma_2 = - \mathrm{i}  \sigma_2
\end{align}
here $\sigma$'s are the Pauli matrices. With $\phi_\alpha=\psi$, $A_\lambda$ and the Lagrangian density (\ref{lagrange}), the functional integral (\ref{eq:1}) becomes the grand partition function 
of spinor $\hbox{QED}_{2+1}$ at a temperature $T = 1/ \beta$  with the free fermion propagator in energy-momentum space given by:
\begin{align} 
  S_F (p) = \frac{\mathrm{i}}{(\mathrm{i} \lambda + \mu) \sigma_3 - \mathrm{i} \vec{\sigma} \cdot \vec{p}}  
\label{eq:2}
\end{align}
where $\vec \sigma = (\sigma_1 , \sigma_2)$ and the Matsubara energy $\lambda = 2 \pi T(n+\frac{1}{2}), \quad n \in \mathbb{Z}$.

\begin{figure}[!htb]
  \centering{\includegraphics[width=6cm]{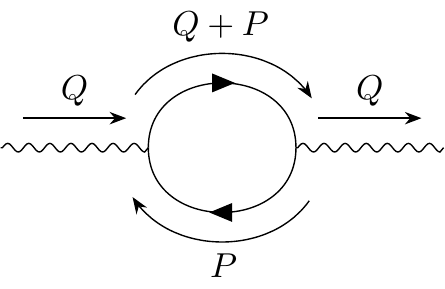}}
  \caption{ the one-loop diagram in spinor QED}
  \label{fig:fermion-loop}
\end{figure}

Our interest in this paper lies on the electric component, i.e.$tt$-component 
of the polarization tensor, $\Pi_{\mu\nu}(q)$, corresponding to the one-loop Feynman diagram in Fig. \ref{fig:fermion-loop}

We have 
\begin{align}
  \Pi_{tt} (q)\equiv e^2 \sigma^\mathrm{sc}(q)
\end{align}
where the reduced polarization tensor $\sigma^\mathrm{sc}(q)$ is given by
\begin{align}
  \sigma^\mathrm{sc} (q) = \frac{1}{\beta} \sum_{n=-\infty}^{+\infty} \int \frac{\mathrm{d}^2 \vec p}{(2\pi)^2} \mathrm{Tr}\,  \bigg[ \sigma_3 S_F(P) \sigma_3 S_F(P+Q) \bigg]
\label{eq:3}
\end{align}
with $P = (\mathrm{i} \lambda , \vec p) ; \, Q=(0,  \vec q)$ and $\vec q =(q,0)$. To explore the Friedel oscillations, we have taken the static limit and aligned the spatial
momentum along x-directions.  
The  Matsubara frequency sum in (\ref{eq:3}) can be converted into a contour integral, and we obtain:
\begin{equation}
\begin{aligned}
  \sigma^\mathrm{sc} (q) &= -2 \int \frac{\mathrm{d}^2 \vec p}{(2 \pi)^2} \oint_{\mathcal{C}_1} \frac{\mathrm{d} z}{2 \pi \mathrm{i}} \; \frac{1}{e^{\beta z } +1} \frac{(z+\mu)^2 + (p_1 + q) p_1 + p_2^2}{ \big[(z+\mu)^2 + |\vec p + \vec q|^2 \big]\, \big[(z + \mu)^2 - p^2\big]}  \\
  &= - \int \frac{\mathrm{d}^2 \vec p}{(2 \pi)^2} \, \bigg\{ \left[\frac{1}{e^{\beta(| \vec p + \vec q| -\mu)} +1}-\frac{1}{e^{-\beta(|\vec p + \vec q| +\mu)} +1}\right]
  \frac{(\vec p + \vec q)^2 + (p_1 + q) p_1 + p_2^2}{|\vec p + \vec q| \big[ (\vec p + \vec q)^2 - \vec p^2 \big]}     \\
  & \hspace{2.7cm}- \left[\frac{1}{ e^{\beta (p - \mu)} + 1}-\frac{1}{ e^{-\beta (p + \mu)} + 1}\right] \frac{p^2 + (p_1 + q)p_1 + p_2^2}{p \big[ p^2 - (\vec p + \vec q)^2 \big]} \bigg\}    \\
  &= \sigma_\text{vac}^\text{sc} (q) +  \sigma_\text{matt}^\text{sc} (q) 
\label{eq:4}
\end{aligned}
\end{equation}
where   $\sigma_\text{vac}^\text{sc}(q) = \sigma^\text{sc}(q)\, \bigg|_{T=0; \, \mu=0}$ and

  \begin{align}
    \sigma_\text{matt}^\text{sc}(q)   &= - \int \frac{\mathrm{d}^2 \vec p}{(2 \pi)^2} \, \bigg\{ \frac{1}{e^{\beta(| \vec p + \vec q| -\mu)} +1} \frac{(\vec p + \vec q)^2 + (p_1 + q) p_1 +p_2^2}{|\vec p + \vec q| \big[ (\vec p + \vec q)^2 - \vec p^2 \big]}    \label{eq:5}  \notag  \\
    & \hspace{2.7cm}- \frac{1}{ e^{\beta (p - \mu)} + 1} \frac{p^2 + (p_1 + q)p_1 - p_2^2}{p \big[ (\vec p + \vec q)^2 - p^2\big]} \bigg\}  + (\mu \leftrightarrow -\mu)  \\
    &= - \mathrm{Re} \; \int  \frac{\mathrm{d}^2 \vec p}{(2 \pi)^2} \, \bigg\{ \frac{1}{e^{\beta(p - \mu)} + 1} \times \frac{1}{p} \times    \notag  \\
    & \hspace{1.2cm}  \left[ \frac{p^2 + p_1(p_1 - q) - p_2^2}{p^2 - (\vec p - \vec q)^2 + \mathrm{i}0^+}  - \frac{p^2 + (p_1 + q) p_1 - p_2^2}{(\vec p + \vec q)^2 - p^2 + \mathrm{i}0^+}\right]  + (\mu \leftrightarrow -\mu) \bigg\}    \label{eq:6} \\
    &=  \frac{1}{2 \pi} \int_0^\infty \ \mathrm{d}p \; \bigg[ 1 + \frac{4 p^2 - q^2}{q \sqrt{q^2 - 4 p^2}} \, \theta(q - 2 p) \bigg] \left( \frac{1}{e^{\beta(p - \mu)} +1} + \frac{1}{e^{\beta(p + \mu)} + 1} \right)   \label{eq:7}
  \end{align}
 In passing, we have added an infinitesimal imaginary part to the two factors $(\vec p + \vec q)^2 - p^2$ in the denominators of (\ref{eq:6}) in order to make each term 
convergent without modifying the real part of the result. The dependence of the distribution function on $\vec q$ is thereby removed by shifting the loop momentum.  Isolating out the finite part , we  obtain \footnote{Here we employ the identity $\left( \frac{1}{e^{\beta(p - \mu)} +1} + \frac{1}{e^{\beta(p + \mu)} + 1} \right) = T\, \sum\limits_{\lambda} \left( \frac{1}{\mathrm{i} \lambda + \mu - p} - \frac{1}{\mathrm{i} \lambda + \mu + p} \right) + 1$, for its proof, one can consult Appendix A of the paper \cite{Yin2016} for its counterpart. }:
\begin{align}
  \sigma_\text{matt}^\text{sc}(q)   &= - \frac{1}{2 \pi q} \int_0^{\frac{q}{2}} \ \mathrm{d}p \sqrt{q^2 - 4 p^2} \bigg[T\, \sum\limits_{\lambda'} \left( \frac{1}{\mathrm{i} \lambda + \mu - p} - \frac{1}{\mathrm{i} \lambda + \mu + p} \right) \bigg] - \frac{q}{16} + R(\mu, T)
  \label{eq:8}
\end{align}
here $R(\mu, T)$ stands for all of the terms independent of the momentum $q$. Taking the derivative with respect to $q$, we obtain that
\begin{align}
  \frac{\mathrm{d}}{\mathrm{d q}} \sigma_\text{matt}^\text{sc}(q)   &= - \frac{1}{2 \pi} \int_0^{\frac{q}{2}} \ \mathrm{d}p \frac{1}{\sqrt{q^2 - 4 p^2}} \bigg[T\, \sum\limits_{\lambda'} \left( \frac{1}{\mathrm{i} \lambda + \mu - p} - \frac{1}{\mathrm{i} \lambda + \mu + p} \right) \bigg] - \frac{1}{16}
  \label{eq:8.1}
\end{align}

Defining a small complex quantity $\epsilon$:
\begin{align}
  \frac{q}{2} =  \mu + \mathrm{i}\lambda    + \epsilon   
\label{eq:9}
\end{align}
substituting (\ref{eq:9}) into (\ref{eq:8.1}) and focusing on term $\lambda'=\lambda$, which makes the integral divergent in the limit $\epsilon\to 0$, 
After evaluating the integral over $p$, the final result reads:

\begin{align}
  \frac{\mathrm{d}  }{\mathrm{d} q} \sigma_\text{matt}^\text{sc}  = \frac{1}{2} \frac{\mathrm{d}}{\mathrm{d} \epsilon} \sigma_\text{matt}^\text{sc} = \frac{\mathrm{i} \, T}{4 \sqrt{\epsilon(q - \epsilon)}} + \text{non-singular terms}
\label{eq:10}
\end{align}
we find that the derivative of $\sigma_\text{matt}^\text{sc}$ diverges as $O\left(\dfrac{1}{\sqrt{\epsilon}} \right)$ in the limit $\epsilon \to 0$.  Besides, if starting with the definition :
\begin{align}
  \frac{q}{2} =  - [\mu + \mathrm{i}\lambda]    + \epsilon
\end{align}
we will obtain the same conclusion. Therefore, the function $\sigma^\text{sc}(q)$ shows square root singularities at 
\begin{align}
  q = \pm 2 (\mu + \mathrm{i} \lambda) \label{eq:108}
\end{align}
on the complex $q$-plane.

\subsection{ Boson case: Scalar QED }
\label{sec:boson-case:-scalar}

In the above subsection,  we have obtained all singularities of photon self-energy in spinor QED$_{2+1}$ with both nonzero temperature and nonzero chemical potential 
at the one-loop level. As a comparison, we will investigate its antithesis; the charged scalar QED$_{2+1}$. In this case, the Lagrangian  
in (\ref{eq:1}) becomes
\begin{align}
  \mathcal{L}[\phi, \phi^\dagger ] = - \big( D^\mu \phi )^\dagger D_\mu \phi - m^2 \phi^\dagger \phi   \label{eq:102}
\end{align}
where the co-variant derivative operator is   $D_\mu \equiv \partial_\mu - \mathrm{i} e A_\mu$. A nonzero mass term is included in order for the relativistic Bose distribution 
function to be well-defined. The free boson propagator of 4-momentum $(\nu_n,\vec k)$  takes the form
\begin{align}
  \Delta_F( \mathrm{i} \nu_n , \vec{k}) = \frac{1}{- (\mathrm{i} \nu_n  +  \mu)^2 + |\vec{k}|^2 + m^2 } 
\end{align}
where the Matsubara frequency is $\nu_n \equiv  2 \pi n T$ with $n \in \mathbb{Z}$ and $\mu$ is the chemical potential for boson system. 

\begin{figure}[!htb]
  \centering{\includegraphics[width=6cm]{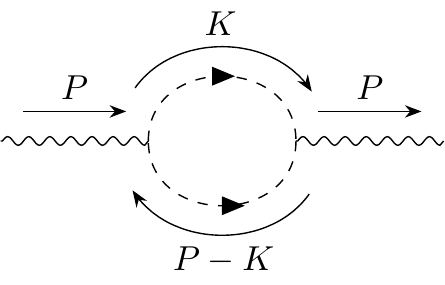}}
  \caption{ the one-loop diagram in scalar QED}
  \label{fig:scalar-loop}
\end{figure}

The one-loop approximation of the polarization function in  scalar QED involves two diagrams, for our purpose,  we're only interested in the one that the momentum of photon has a contribution to momentum integral , hence the polarization function (\ref{eq:3}) corresponds to the Feynman diagram in Fig~\ref{fig:scalar-loop}.  and is given by
\begin{equation}
\begin{aligned}
  \sigma^\text{sc} (q) &= - \frac{1}{\beta}  \sum_{s= -\infty}^{+\infty} \int \frac{\mathrm{d}^2 \vec{k}}{(2 \pi)^2} \,  \Delta_F(\mathrm{i} \nu_n + \mathrm{i} \nu_s, k + q) \Delta_F( \mathrm{i} \nu_s , \vec{k}) \big[ \mathrm{i} \nu_s + \mu + 2 \mathrm{i} \nu_n \big]^2 \\
  &= \sigma^\text{sc}_\text{vac}(q) +  \sigma^\text{sc}_\text{matt}(q)  \label{eq:101}
\end{aligned}
\end{equation}
where $ \sigma^\text{sc}_\text{vac}(q) \equiv  \sigma^\text{sc}(q)\bigg|_{T=0;\, \mu=0}$ and the static limit is taken with the spatial photon momentum $q =|\vec q |$.
Calculating the  Matsubara frequency sum with the aid of a contour integral and shifting the loop momentum appropriately, we find that the last part in (\ref{eq:101}) is given by

\begin{align}
  \sigma^\text{sc}_\text{matt}(q) &= \frac{1}{2} \int \frac{\mathrm{d}^2 \vec{l}}{(2 \pi)^2 } \frac{\sqrt{l^2 + m^2}}{\mathrm{e}^{ \beta (\sqrt{l^2 + m^2} - \mu)} - 1} \left[ \frac{1}{|\vec{l}|^2 - |\vec{l} - \vec{q} |^2} - \frac{1}{|\vec{l} + \vec{q} |^2 - |\vec{l}|^2 } \right] +  \big(\mu \to - \mu \big)  \notag \\
  &= - \frac{1}{q} \int_0^{\frac{k}{2}} \frac{ l \cdot \sqrt{l^2 + m^2}}{\sqrt{q^2 - 4 l^2 }} \, \frac{\mathrm{d}l}{2\pi} \left[ \frac{1}{\mathrm{e}^{\beta (\sqrt{l^2 + m^2} - \mu)} - 1} + \frac{1}{\mathrm{e}^{\beta (\sqrt{l^2 + m^2} + \mu)} - 1} \right]  \label{eq:103} \\
 &= \frac{T}{4 \pi q} \sum_{n' = -\infty}^{+\infty} \int_0^{\frac{q}{2}} \ \mathrm{d}l \frac{ l \sqrt{l^2 + m^2}}{\sqrt{\frac{q^2}{4} - l^2}} \bigg( \frac{1}{\mathrm{i} \nu_{n'} + \mu - \sqrt{l^2 + m^2}}  \notag \\
 & \hspace{6cm} - \frac{1}{\mathrm{i} \nu_{n'} + \mu + \sqrt{l^2 +m^2}} \bigg) + R(\mu, T)  \label{eq:107}
\end{align}
here $R(\mu , T)$  refers to the terms independent of the external momentum $q$.  Introducing a new integration variable $x \equiv \frac{\sqrt{l^2 + m^2}}{s_0}$ and new parameter,  
$s_0 \equiv \sqrt{m^2 + \left( \frac{q}{2}\right)^2}$, we have $\left( \frac{q}{2}\right)^2 - l^2 = s_0^2 \left[1- x^2 \right]$ hence
\begin{align}
  \sigma_\text{matt}^\text{sc}(q) &= \frac{T}{4 \pi q} \sum_{n' = -\infty}^{+\infty} \int_{m/s_0}^{1} \ \mathrm{d} x \frac{x^2}{\sqrt{1 - x^2}} \left( \frac{1}{\mathrm{i} \nu_{n'} + \mu - s_0 x} - \frac{1}{\mathrm{i} \nu_{n'} + \mu + s_0 x}\right)  \label{eq:104}
\end{align}

The singularity occurs when $s_0$ approaches to one of $i\nu_n+\mu$, or $-i\nu_n-\mu$. Denoting $s_0=i\nu+\mu+\epsilon$ with $\nu$ one of $\nu_n$ and $z_0 \equiv \frac{\mathrm{i} \nu + \mu}{\mathrm{i}\nu + \mu + \epsilon}$,  
and isolating out the diverging term $n'=n$ in (\ref{eq:104}) from others in the limit $\epsilon\to 0$, we end up with
\begin{align}
  \sigma_\text{matt}^\text{sc}(q) &= \frac{T}{4 \pi q} \frac{1}{s_0} \int_{\frac{m}{s_0}}^1 \ \mathrm{d}x \frac{1}{\sqrt{1 - x^2}} \left[ \frac{1}{z_0 - x} - \frac{1}{z_0 + x} \right] + \text{non-singular terms} \\
 &= - \frac{\mathrm{i}T}{4 q} \frac{1}{\sqrt{2 ( \mathrm{i} \nu + \mu) \epsilon}} + \text{non-singular terms} \label{eq:105}
\end{align}
which indicates the divergent behavior 
\begin{align}
  \sigma^\text{sc} (q)\sim O\left(\frac{1}{\sqrt{\epsilon}} \right) , \quad \text{as} \quad  \epsilon \to 0  
\end{align}
and the location of the singularities on the complex $q$-plane 
\begin{align}
  q = \pm 2 \sqrt{(\mathrm{i} \nu + \mu)^2 - m^2}  \label{eq:106}
\end{align}

From the previous results,  we see that the locations of the singularities of the electric polarization function of spinor or scalar QED3 on the complex momentum plane stem from 
the poles of the fermionic or bosonic distribution functions and all the singularities are the branch points of square root.
A difference between bosonic and fermionic cases is that we have to introduce a mass for boson with $|m|< |\mu|$ to make the integration of 
Bose-Einstein distribution $n_B(E) = 1/\big(\mathrm{e}^{\beta(E \pm \mu)} - 1\big)$ well defined, while for the fermionic case, the mass is optional. Asymptotically, the singularities in either case are distributed along two straight lines $q=\pm\mu$ 
with equal spacing for large magnitude of $q$, independent of the mass. As we shall see, a similar asymptotic distribution emerges in strong coupling.

\section{Static Electric Polarizability in  Gauge/Gravity   Duality}
\label{sec-3}

The prototype of the gauge/gravity duality is the correspondence between the $\mathcal{N}=4$ $SU(N_c)$ super Yang-Mills and the type IIB superstring in AdS$_5\times$S$_5$ with the former residing 
on the boundary of AdS$_5$. In particular, in the limit of 
large $N_c$ and strong 't Hooft coupling, the superstring side of the correspondence reduces to a classical supergravity and becomes tractable.
Motivated by this relationship, the bottom-up approach to the gauge/gravity duality amounts to identify the classical action of the gravity-matter system 
in the bulk geometry of an asymptotically AdS boundary with 
the quantum effective action of some strongly interacting field theory on the boundary, i.e. 
\begin{align}
  \Gamma[\mathring{\phi}_\alpha] = S_\mathrm{cl.}[\mathring{\phi}_\alpha]    \label{eq:11}
\end{align}
where $\mathring{\phi}_\alpha$ represent the boundary value of the solution to the bulk equation of motion for the matter fields or metric tensor. 
Applying the linear response theory, we can obtain the one-particle irreducible correlator of the strongly-coupled system:
\begin{align}
  \langle \mathcal{O}_1(x) \mathcal{O}_2(x) \rangle_\text{1PI} = \frac{\delta^2 S_\mathrm{cl.}}{\delta \mathring{\phi}_i(x_1) \mathring{\phi}_i(x_2)} \bigg|_{\mathring{\phi}_j = 0} , \qquad j = 1,2 , \cdots, \text{but } \; j \ne i
\label{eq:12}
\end{align}
with the operator $\mathcal{O}$ dual to $\mathring{\phi}_\alpha$.

\subsection{Gravity Preliminaries in the Bulk Theory}
\label{sec:grav-prel-bulk}
\label{sec-3.1}

In order to generate the strong coupling counterpart of the polarization tensor discussed in section \ref{sec:one-loop-correction} via gauge/gravity duality, 
we consider an Einstein-Maxwell theory in the bulk with a Reissner-Nordstr\"om black hole and an asymptotically $AdS_4$ boundary. The action on RHS of (\ref{eq:11}) takes the form
\begin{align}
  S &= \int \ \mathrm{d}^4 x \, \sqrt{-g}\; \bigg\{   G_4\, \big( R - 2 \Lambda \big)-  K_4\, \big( F_{\mu \nu} F^{\mu \nu} \big)  \bigg\} 
\label{eq:13}
\end{align}
where $R$ is the scalar curvature, $\Lambda$ is the negative cosmological constant, $\Lambda = -\frac{3}{L^2}$,  $L$ is the AdS radius, 
and $F_{\mu\nu}$ is a $2$-form field ,$F_{\mu\nu} = \partial_\mu A_\nu - \partial_\nu A_\mu$. For the clarity of notations, we shall scale 
the gauge potential such that  $\frac{K_4}{G_4} = L^2$. 

The action (\ref{eq:13}) leads to a solution of equations of motion with a RN-AdS black hole metric and a background gauge potential and its on-shell value corresponds to the thermodynamic potential of the 
normal phase of a holographic superconductor where the charged scalar field in the bulk vanishes \cite{Gubser2008, Hartnoll2008,Hartnoll2008a, Horowitz2008, Yin2015}. 
In terms of the Poincar\'{e} frame, the background black hole metric reads
\begin{align}
  \mathrm{d}\bar s^2 = \bar g_{\mu \nu} \ \mathrm{d}x^\mu \ \mathrm{d}x^\nu = \frac{L^2}{z^2} \left( -f(z) \ \mathrm{d}t^2 + \frac{\mathrm{d}z^2}{f(z)} 
    + \ \mathrm{d}x^2 + \ \mathrm{d}y^2 \right) 
\label{eq:14}
\end{align}
where the metric function 
\begin{align}
  f(z) = 1 - (1+Q^2) \, \left(\frac{z}{z_+} \right)^3 +  Q^2 \, \left(\frac{z}{z_+} \right)^4 
\label{eq:15}
\end{align}
and the background gauge potential 
\begin{align}
  \bar A = \bar A_t \ \mathrm{d}t = \mu \left( 1 - \frac{z}{z_+}\right) \ \mathrm{d}t 
\label{eq:16}
\end{align}
where  $z_+$ refers to the horizon in this coordinates, $Q$ is the charge of the black hole and $\mu$ corresponds to the chemical potential of the boundary field theory.
The  Hawking temperature of the RN-AdS  black hole  is given by
\begin{align}
  T = \frac{1}{4 \pi z_+}\left(  3 - Q^2 \right)=\frac{\mu}{Q}\frac{3-Q^2}{ 4 \pi} 
\label{eq:17}
\end{align}
which is also the temperature of the boundary field theory in equilibrium and the charge is related to the chemical potential via $Q = \mu z_+$ 

Because of (\ref{eq:17}), the positivity of the temperature requires $Q^2 \in (0,3)$, which makes $z=z_+$ the closest zero to the AdS boundary $z=0$ for real $z$ and thereby the horizon of the RN-AdS black hole. 
The physical domain of the radial coordinate is therefore $0\le z\le z_+$. 

\subsection{Fluctuations of gauge fields and metric fields}
\label{sec:fluct-gauge-fields}

According to the holographic dictionary, the electric current operator and the energy-momentum tensor on the boundary field theory are dual to the fluctuations of
 the gauge fields and metric tensor in the bulk theory, respectively. The static electric Green function in the strong coupling limit we are focusing on will be 
obtained from the solution of the bulk equations of motion for these fluctuations.

Starting with the definitions
\begin{align}
  g_{\mu \nu} &= \bar g_{\mu \nu} + h_{\mu \nu}   \\
  A_\mu  &= \bar A_\mu + a_\mu 
\label{eq:18}
\end{align}
where the background fields ($\bar g_{\mu \nu} , \bar A_\mu$) refer to the background solutions (\ref{eq:14}) and (\ref{eq:16}), and ($h_{\mu \nu}, a_\mu$) 
represent the corresponding fluctuations, respectively.
\footnote{To maintain  the basic property of a metric  tensor $ g_{\mu \rho} g^{\rho \nu}=\delta_\mu^\nu $ , we have $g^{\mu \nu} = \bar g^{\mu \nu} - h^{\mu\nu} , 
\quad  \text{and} \quad \sqrt{-g} = \sqrt{ -\bar g } \big( 1 + \frac{1}{2}\bar g^{\mu \nu} h_{\mu \nu} + O\big( h^2 \big) \big)$ .} 
The two-point Green's functions are extracted from the quadratic terms of the on-shell bulk action in the boundary values of the fluctuations 
($h_{\mu \nu}, a_\mu$), thus we need only to solve the Einstein-Maxwell equations up to the 1st-order in $h_{\mu\nu}$ and $a_\mu$. We work in the radial gauge 
\begin{align}
  h_{z \nu} = 0 , \qquad a_z = 0  , \quad \nu = \{t,x,y,x\}
\label{eq:19}
\end{align}
and in frequency-momentum space
\begin{equation}
  \begin{aligned}
    h_{\mu \nu} (t, z, x ,y) &\sim e^{ \mathrm{i}(-\omega t + q x) } \, h_{\mu \nu} (z | \omega, q)  \\
    a_\mu(t, z , x, y) &\sim e^{\mathrm{i}(-\omega t + q x)} \, a_\mu(z | \omega, q)   
\label{eq:20}
  \end{aligned}
\end{equation}
where the fluctuations turn into the functions of the radial variable $z$,  energy $\omega$ and momentum $q$. We have aligned the momentum along the x-axis by taking advantages of the $SO(2)$ symmetry in $x-y$ plane, such that the fluctuation fields are classified into two categories according to the parity under $y \to -y$
\begin{align}
  \text{Odd Parity:} &\quad h_{ty} , \; h_{xy} , \;  a_y \\
  \text{Even Parity:} &\quad h_{tt} , \; h_{tx} , \;  h_{xx} , \; h_{yy} ,\; a_t ,\; a_x 
\end{align}
In \cite{Yin2016}, we have studied the fluctuation underlying the transverse polarization, $a_y$, which belongs to the odd parity group. The fluctuation underlying the electric polarization we are interested here, 
$a_0$ belongs to the even parity group.  As we shall see, in the static limit, $a_x$ and $h_{tx}$ are decoupled from the other components of even parity \cite{Edalati2010c}. 

In the static limit, $\omega \to 0$,  the linearized Einstein-Maxwell for $h_\mu(z , q)$ and $a_t(z,q)$ read
\begin{align}
  0 &=     {h_{\, t}^t}'' +  \frac{3}{2} \frac{u^2 }{f}\left(\frac{ f }{ u^2 } \right)' {h_{\, t}^t}' + \frac{ u^2 }{ 2 f } \left(\frac{ f }{ u^2 } \right)' [{h_{\, x}^x}' + {h_{\, y}^y}'] + \frac{Q^2}{f}\big(2  u^2 - \mathfrak{q}^2 \big) h_{\, t}^t + 4 \frac{ Q^2 }{ \mu } \frac{u^2 }{f} a_t'      \label{eq:21}\\
     0 &=  \big[ {h_{\, t}^t}'' + {h_{\, x}^x}'' + {h_{\, y}^y}'' \big] +\frac{u^2}{2f} \left(\frac{ f }{ u^2 } \right)'  [{h_{\, t}^t}' + {h_{\, x}^x}' + {h_{\, y}^y}'] + \frac{ f' }{ f } {h_{\, t}^t}' \notag \\
 & \hspace{9cm}  + 2 Q^2 \frac{ u^2 }{ f } h_{\, t}^t + 4 \frac{ Q^2 }{ \mu } \frac{ u^2 }{ f } a_t'     \label{eq:22} \\
 0 &= {h_{\, x}^x}'' +  \frac{1}{f} (Q^2 u^3 - \frac{ 3 }{ u }  ) {h_{\, x}^x}' - \frac{ 1 }{ u } [ {h_{\, t}^t}' + {h_{\, y}^y}' ]  -  Q^2 \left(\mathfrak{q}^2 + 2  u^2 \right) h_{\, t}^t  \notag \\
 & \hspace{9cm} - Q^2 \mathfrak{q}^2  h_{\, y}^y - 4 \frac{ Q^2 }{ \mu } u^2  a_t'   \label{eq:23}  \\
   0 &=  {h_{\, y}^y}'' + \frac{1}{f} (Q^2 u^3 - \frac{ 3 }{ u } ) {h_{\, y}^y}' - \frac{ 1 }{ u }  [ {h_{\, t}^t}' + {h_{\, x}^x}' ]  - Q^2 \mathfrak{q}^2 h_{\, y}^y   - 2 Q^2 u^2  h_{\, t}^t- 4 \frac{ Q^2 }{ \mu } u^2  a_t'   \label{eq:24}  \\
   0 &=   \big[ {h_{\, t}^t}' + {h_{\, y}^y}' \big] + \frac{ f' }{ 2 f } h_{\, t}^t + 4 \frac{ Q^2 }{ \mu } u^2 a_t    \label{eq:25} \\
   0 &=  a_t'' - \frac{Q^2 }{f} \mathfrak{q}^2 \, a_t  + \frac{ \mu }{ 2 }  \big( {h_{\, t}^t}' - {h_{\, x}^x}' - {h_{\, y}^y}'\big)   \label{eq:26}  \\    
   0&=   {h_{\, t}^x}'' - \frac{ 2 }{ u } {h_{\, t}^x}'  - 4 \frac{ Q^2 }{ \mu }  \frac{u^2}{f}  a_x'     \label{eq:27}  \\
   0 &=   \mathfrak{q} (f {h_{\, t}^x}' - f' h_{\, t}^x)    \label{eq:28} \\
   0 &=  \big(f a_x' \big)' - \mu  {h_{\, t}^x}'    \label{eq:29}
\end{align}
where we have introduced two dimensionless quantities 
\begin{align}
   \mathfrak{q} = \frac{q}{\mu} , \quad u = \frac{z}{z_+} \label{eq:30}
\end{align}
the prime in those equations means the derivative with respect to $u$, and  the AdS radius $L=1$ is set to one. Then the metric function becomes $f=1-(1+Q^2)u^3+Q^2u^4$.

It's easy to see from (\ref{eq:21}) to (\ref{eq:29}) that in the static limit, the even parity is splitted into two independent subsets: 
\begin{align}
\{h_t^t , h_x^x, h_y^y, a_t \} \quad \text{and}  \quad \{ h_t^x , a_x \}
\end{align}
and our interest lies in the first one because  the electric polarization is extracted from the fluctuation $a_t$ evaluated on the boundary. To disentangle the coupling between $a_t$ and metric fluctuations
 $h_{\, t}^t , h_{\, x}^x$ and $h_{\, y}^y$ in (\ref{eq:21})-(\ref{eq:26}) we employ the master fields method \cite{Edalati2010c, Kodama2003a, Kodama2004} that converts those coupled equations into  
two decoupled master field equations.
The two gauge invariant master fields, $\Phi_\pm(u | \mathfrak{q})$, are defined by
\begin{align}
  \Phi_\pm  \equiv  \alpha_\pm \, \Phi_1 + \frac{Q^2}{\mu} \Phi_2    \label{eq:31}
\end{align}
and satisfy the following two ordinary differential equations:
\begin{align}
  \Phi_\pm'' + \frac{f'}{f} \Phi_\pm'  - \frac{U_\pm}{f^2} \, \Phi_\pm = 0   \label{eq:32}
\end{align}
where
\begin{align}
  \Phi_1 &\equiv \frac{1}{u}\, h_{\; y}^y + \frac{f}{ Q^2[k^2 - Z^2] \, u^2 - f' \, u } \, \left( {h_{\; x}^x}' + {h_{\; y}^y}' \right)  \label{eq:94} \\
 \Phi_2 &\equiv  - a_t' + \mu \, h_{\, y}^y - \frac{\mu}{2} \, h_{\, t}^t   \label{eq:100}
\end{align}
and both $\alpha_\pm$ and $U_\pm$ are the functions of $u, \mathfrak{q}$. We left the detailed forms of $\alpha_\pm$ and $U_\pm$ to the appendix \ref{sec:mast-field-equat}.

The master field equations (\ref{eq:32}) can be transformed into a Schr\"{o}dinger-like equations 
\begin{align}
  \Psi_\pm''(u | \mathfrak{q})  + V_\pm(u | \mathfrak{q}) \, \Psi_\pm(u | \mathfrak{q})  = 0   \label{eq:34}    
\end{align}
via the definition 
\begin{align}
  \Phi_\pm(u | \mathfrak{q}) \equiv \frac{1}{\sqrt{f(u)}} \, \Psi_\pm(u | \mathfrak{q})   \label{eq:33}
\end{align}
and 
\begin{align}
  V_\pm(u | \mathfrak{q}) =  V_\pm(u | k) = - \bigg[  \frac{U_\pm}{f^2} + \frac{1}{4} \left(\frac{f'}{f}\right)^2 + \frac{1}{2} \left(\frac{f'}{f}\right)'  \bigg]   
\label{eq:93}
\end{align}
where we define the modified momentum $k$ as
\begin{align}
  k^2 \equiv \mathfrak{q}^2 + \left[ \frac{3}{4}\left( 1 + \frac{1}{Q^2}\right)\right]^2
\label{eq:37}
\end{align}

According to the correspondence (\ref{eq:11}) between the quantum effective action of the boundary field theory and the bulk action, the on-shell action (\ref{eq:13}) corresponding to 
the linearized Einstein-Maxwell equations is a quadratic functional 
of the boundary values of all $h_{\, \mu}^\nu$ and $a_\mu$ with the coefficients corresponding to various two point 1PI Green's functions in strong coupling limit.  We denote $\mathcal{C}_{tt}$ 
as the electric polarization, $\mathcal{C}_{tt} (\omega =0, \mathfrak{q})= \sigma^\mathrm{sc}(\mathfrak{q})$ . After imposing the asymptotic conditions
\begin{align}
 \lim_{ u \to 0} a_x \equiv \mathring{a}_x =0 , \quad \lim_{u \to 0} a_y \equiv  \mathring{a}_y =0 \quad \text{ and } \lim_{u \to 0} h_{\mu\nu} \equiv  \mathring{h}_{\mu\nu} = 0  
\label{eq:35}
\end{align}
$\mathcal{C}_{tt}$, extracted from the functional derivative  in (\ref{eq:12}), reads 
\begin{equation}
  \begin{aligned}
  \mathcal{C}_{tt}(\omega, \mathfrak{q}) &=  K_4 \, \lim_{u \to 0} \sqrt{- \bar g} \, \bar g^{uu} \bar g^{tt} \; \frac{a_t'}{a_t}      \\
  &=- \frac{K_4}{z_+} \, \lim_{u \to 0}  \frac{a_t'}{a_t} 
\label{eq:36}
\end{aligned}
\end{equation}
Owing to (\ref{eq:31}) and (\ref{eq:33}), we obtain 
\begin{align}
  a_t'(u | k) \equiv  \frac{1}{\sqrt{f}} \, \frac{\mu}{2 Q^2 k} \bigg\{  (Z-2u) [ \Psi_+ - \Psi_-] - k [\Psi_+ + \Psi_-] \bigg\} + \mu \, h_{\; y}^y - \frac{\mu}{2}\, h_{\; t}^t 
\label{eq:38}
\end{align}
where the dimensionless quantity
\begin{align}
  Z \equiv  \frac{3}{4}\left(1 + \frac{1}{Q^2} \right)
\label{eq:39}
\end{align}

The situation to tackle the solution for (\ref{eq:36}) is different from the case in odd parity \cite{Yin2016} since the equation (\ref{eq:38}) connects the master field to the  first derivative of gauge fluctuation. 

\subsection{ Analyticity of the electric polarization  with respect to the momentum }
\label{sec:analyt-electr-polar}

    Unlike the odd parity case, the master field equations are far more complicated here. Also the intricate links from the master field to $a_t$, shown in  (\ref{eq:34}) , (\ref{eq:36}) and (\ref{eq:38}), 
prohibit a straightforward way to read off the analyticity
 of $\mathcal{C}_{tt}(0 | \mathfrak{q} )$ with respect to  the momentum $q$ from the master field equations. 
The alternative strategy we are following is to investigate the coupled Einstein-Maxwell equations directly. Our main concern is the analytic property of the fluctuation $a_t(u|q)$ 
and its derivative $a_t^\prime(u|q)$ in the limit $u\to 0$. We shall prove below that $\mathcal{C}_{tt}(0, \mathfrak{q} )$ is a meromorphic function of $q$.

It is convenient to work with the new variable $ \zeta \equiv 1- u$ , hence $\zeta = 0$ represents the horizon, and the prime in this section refers to the derivative with respect to $\zeta$, 
e.g. $a_t'\equiv \frac{\mathrm{d}a_t}{\mathrm{d}\zeta}$. Following Ref.\cite{Blake2015}, equations (\ref{eq:21})(\ref{eq:22})(\ref{eq:23}) and (\ref{eq:24}) can be combined to cancel the 
2nd order derivatives and to yield an expression for ${h_{\; x}^x}^\prime$
\begin{align}
  {h_{\; x}^x}^\prime &= -{h_{\; y}^y}^\prime + \frac{2u}{f' + 4 f} \bigg\{ - \frac{2 f}{u} {h_{\; t}^t}^\prime +  Q^2\big[ \mathfrak{q}^2 + 2 u^2   \big] h_{\; t}^t +  Q^2 \mathfrak{q}^2 h_{\; y}^y  - 4  \frac{Q^2}{\mu}\, u^2 a_t' \bigg\} 
\label{eq:41}
\end{align}
Substituting (\ref{eq:41}) into (\ref{eq:24}) and (\ref{eq:26}), we are left with the three coupled Einstein-Maxwell equations (\ref{eq:24})(\ref{eq:25}) and (\ref{eq:26}) for $a_t$, $h_{\; t}^t$ and $h_{\; y}^y$ ,respectively
\begin{equation}
\begin{aligned}
 0 &= {h_{\; y}^y}^{\prime\prime} + \frac{C_3}{\mu \zeta}\,  a_t^\prime +C_4 {h^t_t}^\prime + \frac{C_5}{\zeta}\,  {h_{\; y}^y}^\prime + \frac{D_6}{\zeta}  h_{\; t}^t + \frac{D_7}{\zeta} h_{\; y}^y \\
 0 &= {h_{\; t}^t}^\prime + {h_{\; y}^y}^\prime + \frac{D_1}{\mu \zeta}\,  a_t + \frac{D_2}{\zeta}\, h^t_{\; t}   \\
 0 &= a_t^{\prime\prime} + C_1 a_t^\prime + \mu C_2 \,  {h^t_{\; t}}^\prime + \frac{D_3}{\zeta}\, a_t + \mu D_4\, h^t_t + \mu D_5\, h_{\; y}^y  
\label{eq:44}
\end{aligned}
\end{equation}
where the coefficients are given by 
\begin{equation}
  \begin{aligned}
    C_1 &=  - \frac{4 Q^2 u^2}{f'+\frac{4f}{u}}  &\quad   
    C_2 &= - \frac{1}{2}\frac{ u f'+8f }{ u f' + 4f } \\
    C_3 &= \frac{4Q^2\zeta u^2}{ f}\frac{f'+\frac{2f}{u}}{f'+\frac{4f}{u}}  &\quad  
    C_4 &= \frac{ f' }{ u f' + 4f } \\
    C_5 &= \left(\frac{f'}{f} + \frac{2}{u}\right)\zeta   &  & \\
    D_1 &=  -\frac{4Q^2\zeta}{ f} u^2  &\quad 
    D_2 &= \frac{\zeta f'}{2f}  \\
    D_3 &= -\frac{Q^2 \mathfrak{q}^2}{f}\zeta  &\quad 
    D_4 &= \frac{ Q^2 \, u (2u^2  + \mathfrak{q}^2)}{ u f' + 4f }  \\
    D_5 &= \frac{ Q^2 \mathfrak{q}^2 \, u }{ u f' + 4f }  & \quad 
    D_6 &= - \zeta \, \frac{ 2 Q^2 }{ f }\frac{ u^2\left( u f' + 2f\right) -f \mathfrak{q}^2 }{ u f' + 4 f  }  \\
    D_7 &= - \zeta Q^2 \frac{ \mathfrak{q}^2 }{ f }  \frac{ uf'+ 2 f }{ u f'+ 4 f }   & & 
\label{eq:45}
  \end{aligned}
\end{equation}
and they are all analytic and nonvanishing at the horizon $\zeta =0$. Making Taylor expansions, we have
\begin{align}
  C_i &= \sum\limits_{m=0}^\infty c_i^{(m)} \, \zeta^m   \label{eq:42}\\
  D_i &= \sum\limits_{m=0}^\infty  d_i^{(m)} \, \zeta^m  \label{eq:43}
\end{align}
where all coefficients are polynomials in momentum $q$. 

Now we are ready to explore the solution that is analytic at the horizon. 
It follows from the third equation of 
(\ref{eq:44}) that $a_t=O(\zeta)$ in the limit $\zeta\to 0$. Then the second equation of (\ref{eq:44}) implies that $h^t_t=O(\zeta)$ in the limit $\zeta\to 0$
as well. Therefore the power series solution of (\ref{eq:44}) takes the form
\begin{align}
  a_t &= \mu \, \zeta \sum_{n=0}^\infty   \alpha_n \zeta^n \equiv a_t(u|\mathfrak{q})  \label{eq:49}  \\
  h^t_t &= \sum_{n=0}^\infty \beta_n \zeta^n \equiv h^t_t(u|\mathfrak{q})  \label{eq:50}\\
  h^y_y &= \sum_{n=0}^\infty\gamma_n\zeta^n \equiv h^y_y(u|\mathfrak{q})  \label{eq:52}
\end{align}
For the later convenience, we use the same format for the series of $h^t_t$ as that for $h^y_y$ but set $\beta_0=0$.
Substituting  series (\ref{eq:42})-(\ref{eq:52}) into equations (\ref{eq:44}) , we find the following recursion relations:
\begin{equation}
  \begin{aligned}
    (n+1)(n+c_5^{(0)}) \gamma_{n+1} &+ \sum_{k=0}^n(k+1) \, c_3^{(n-k)} \alpha_k + \sum_{k=0}^n \, k \, c_4^{(n-k)} \beta_k + \sum_{k=0}^n \, k \, c_5^{(n-k+1)} \gamma_k   \\
    &+  \sum_{k=0} \, d_6^{(n-k)} \, \beta_k + \sum_{k=0}^n \, d_7^{(n-k)} \, \gamma_k = 0
\label{eq:46}
  \end{aligned}
\end{equation}
\begin{equation}
  (n+1+d_2^{(0)})\beta_{n+1} + (n+1)\gamma_{n+1} + \sum_{k=0}^n d_1^{(n-k)}\alpha_k+\sum_{k=0}^n d_2^{(n+1-k)}\beta_k=0
\label{eq:47}  
\end{equation}
\begin{equation}
  \begin{aligned}
    (n+1)(n+2) \alpha_{n+1} &+ (n+1)c_2^{(0)}\beta_{n+1}+\sum_{k=0}^n(k+1)c_1^{(n-k)} \alpha_k + \sum_{k=0}^n  k \, c_2^{(n+1-k)} \, \beta_k  \\
    &+ \sum_{k=0}^n d_3^{(n-k)}  \, \alpha_k + \sum_{k=0}^n  \, d_4^{(n-k)} \beta_k +  \sum_{k=0}^n  \, d_5^{(n-k)}\, \gamma_k = 0
\label{eq:48}
  \end{aligned}
\end{equation}
Given $\alpha_k$, $\beta_k$ and $\gamma_k$ for $k\le n$, explicit formulas for $\alpha_{n+1}$, $\beta_{n+1}$ and $\gamma_{n+1}$ can be obtained from (\ref{eq:46}), (\ref{eq:47}) and (\ref{eq:48}). 
Indeed, the formula for $\gamma_{n+1}$ follows from (\ref{eq:46}). Substituting it into (\ref{eq:47}) yields  
the formula for $\beta_{n+1}$. With explicit expressions of $\beta_{n+1}$, the formula for $\alpha_{n+1}$ is obtained from (\ref{eq:48}). These recursion formulas are displayed in the 
Appendix B, which show that each order of a series is a polynomial in $q$ and is an analytic function of $q$. Using an inductive method analogous to that used for the 
analyticity of the solution of a second order ordinary differential equation around a canonical singularity \cite{Whittaker1952}, we are able to prove that the series solution of (\ref{eq:49}), (\ref{eq:50}) and (\ref{eq:52}) converges uniformly with respect to a finite $q$ in a circle $|\zeta|<r$ on 
the complex $\zeta$-plane around the horizon $\zeta=0$. It follows from  Weierstrass M-test theorem \cite{Whittaker1952} that the solutions of (\ref{eq:49}), (\ref{eq:50}) and (\ref{eq:52}) also define an analytic function with respect 
to $q$ for a fixed $\zeta$, so does the derivative of the solution with respect to $\zeta$. Because of the absence of singularities of the coefficients $C_i$'s and $D_i$'s in the segment $0<\zeta<1$ ($0<u<1$), the solutions can be analytically 
continuated to the outside of the convergence circle. For example, we may choose a point $\zeta_0$ , such that $0 < \zeta_0 < r$ and its distance, $d$, 
to the nearest singularity of the coefficients satisfies  $d > r - \zeta_0$. Around the point $\zeta_0$, the three Einstein-Maxwell equations (\ref{eq:24}) (\ref{eq:25}) and (\ref{eq:26}) 
can be written as
\begin{equation}
  \begin{aligned}
    0 &= {h^y_y}^{\prime\prime}+{\frac{ \bar C_3 }{ \mu } } a_t^\prime + {\bar C}_4 {h^t_t}^\prime + {\bar C}_5 {h^y_y}^\prime + {\bar D}_6 h^t_t  + {\bar D}_7 h^y_y \\
    0 &= {h_t^t}^\prime + {h_y^y}^\prime + \frac{ {\bar D}_1 }{ \mu }  a_t + {\bar D}_2 h^t_t  \\
    0 &= a_t^{\prime\prime} + {\bar C}_1 a_t^\prime +\mu  {\bar C}_2 {h^t_t}^\prime +{\bar D}_3 a_t + \mu \bar{D}_4 h^t_t + \mu {\bar D}_5 h^y_y
\label{eq:54}
  \end{aligned}
\end{equation}
where all coefficients are analytic for $|\zeta-\zeta_0|<d$. Substituting the ansatz
\begin{equation}
\begin{aligned}
    a_t(u|\mathfrak{q}) &=\sum_{n=0}^\infty{\bar\alpha}_n(\zeta-\zeta_0)^n \\
    h^t_t(u|\mathfrak{q}) &= \sum_{n=0}^\infty{\bar\beta}_n(\zeta-\zeta_0)^n \\
    h^y_y(u|\mathfrak{q}) &= \sum_{n=0}^\infty{\bar\gamma}_n(\zeta-\zeta_0)^n    
\end{aligned}
\end{equation}
into (\ref{eq:54}), one obtains the recursion relations 
\begin{equation}
  \begin{aligned}
    0 &= (n+1)(n+2) \, {\bar\gamma}_{n+2} + \sum_{k=0}^n(k+1) \, {\bar c}_3^{(n-k)} \, {\bar\alpha}_{k+1}+\sum_{k=0}^n(k+1){\bar c}_4^{(n-k)} \, {\bar\beta}_{k+1} \\
    &\hspace{3cm} + \sum_{k=0}^n (k+1) {\bar c}_5^{(n-k)} \, \gamma_{k+1} +  \sum_{k=0}{\bar d}_6^{(n-k)} \, {\bar\beta}_k + \sum_{k=0}^n \, {\bar d}_7^{(n-k)} \, {\bar\gamma}_k
\label{eq:56}
  \end{aligned}
\end{equation}
\begin{equation}
  0 = (n+2)({\bar\gamma}_{n+2} + {\bar\beta}_{n+2}) + \sum_{k=0}^{n+1} {\bar d}_1^{(n-k+1)} \, {\bar\alpha}_k + \sum_{k=0}^{n+1} \, {\bar d}_2^{(n-k+1)} \, {\bar\beta}_k
\label{eq:57}
\end{equation}
\begin{equation}
  \begin{aligned}
    0 &= (n+1)(n+2){\bar\alpha}_{n+2} + \sum_{k=0}^n(k+1){\bar c}_1^{(n-k)}{\bar\alpha}_{k+1}+\sum_{k=0}^n (k+1){\bar c}_2^{(n-k)}{\bar\beta}_{k+1}\\
    & \hspace{3cm} + \sum_{k=0}^n{\bar d}_3^{(n-k)}{\bar \alpha}_k+\sum_{k=0}^n{\bar d}_4^{(n-k)}{\bar\beta}_k + \sum_{k=0}^n{\bar d}_5^{(n-k)} \, {\bar\gamma}_k
  \label{eq:58}
\end{aligned}
\end{equation}
where $\bar{c}_i^{(m)}$'s and $\bar{d}_i^{(m)}$'s stand for the coefficients of the Taylor expansion of $\bar{C}_i$'s and $\bar{D}_i$'s around $\zeta_0$ and are 
polynomials in $q$. With $a_t$, $h^t_t$, $h^y_y$ and their derivatives at $\zeta_0$, provided by the series solution of (\ref{eq:49}), (\ref{eq:50}) and (\ref{eq:52}), 
all coefficients of (\ref{eq:54}) can be determined and are analytic in $q$. It follows that the series solution (\ref{eq:54}) and its derivative, being convergent for 
$|\zeta-\zeta_0|<d$, are analytic functions of $q$ for a fixed $\zeta$. The procedure can be repeated until we reach a series solution whose convergence circle pass 
through the AdS boundary $\zeta=1$, which is another singularity of the coefficients $C_i$'s and $D_i$'s. The power series solution in the neighborhood of the 
AdS boundary
\begin{equation}
\begin{aligned}
    a_t(u|\mathfrak{q}) &= \sum_{n=0}^\infty a_n(1-\zeta)^n=\sum_{n=0}^\infty a_n u^n  \\
    h^t_t(u|\mathfrak{q}) &= \sum_{n=0}^\infty b_n(1-\zeta)^n=\sum_{n=0}^\infty b_n u^n  \\
    h^y_y(u|\mathfrak{q}) &= \sum_{n=0}^\infty c_n(1-\zeta)^n=\sum_{n=0}^\infty c_n u^n    
\label{eq:55}
\end{aligned}
\end{equation}
shows no diverging behavior in itself and its derivative as $\zeta\to 1$. Consequently, the series solution and its derivatives with respect to $u$ remain convergent in the limit $\zeta\to 1$. It follows 
from the Weierstrass  theorem that  
$a_t(0|\mathfrak{q})$, $h^t_t(0|\mathfrak{q})$ and $h^y_y(0|\mathfrak{q})$ together with their derivatives are all analytic in $q$.The details behind each step of the proof outlined above can be found in Appendix B.

A subtlety arises: The series solution of $a_t$ is developed from two arbitrary constants, $\alpha_0$ and $\gamma_0$, which are not sufficient to make $h^y_y(0|\mathfrak{q})$ and $h^t_t(0|\mathfrak{q})$ vanish while 
maintaining a nonzero value of $a_t(0|\mathfrak{q})$. This issue was noted in \cite{donos2014,Blake2015} and the reason is that we take the radial gauge and demand the analyticity of $a_t(u|\mathfrak{q})$, 
$h^y_y(u|\mathfrak{q})$ and $h^t_t(u|\mathfrak{q})$ in $u$ at the horizon.
Unlike the master fields, $a_t(u|\mathfrak{q})$, $h^y_y(u|\mathfrak{q})$ and $h^t_t(u|\mathfrak{q})$ are not gauge invariant, a coordinate transformation
\begin{eqnarray}
  z &\to& z + \mathrm{e}^{iqx}\phi(u)  \notag \\
x &\to& z + \mathrm{e}^{iqx}\chi(u)
\label{transf}
\end{eqnarray}
with
\begin{eqnarray}
\phi(u) &=& Au\sqrt{f(u)} \notag \\ 
\chi(u) &=& - \mathrm{i} q A\int_0^u \ \mathrm{d}\xi \, \frac{\xi}{ \sqrt{f(u)} }
\end{eqnarray}
can be made {\it within} the radial gauge at a cost of introducing a square root singularity $\sim O(\sqrt{1-u})$ at the horizon. Under this transformation, we find that
\begin{eqnarray}
  \delta h^y_y\ \big|_{u=0} &=& \delta h^t_t\ \big|_{u=0} =-2A\nonumber\\
  \delta a_t\ \big|_{u=0} &=& 0  \\
\delta a_t^\prime \ \big|_{u=0} &=& \mu A \notag
\end{eqnarray}
Therefore, we may choose the constants $\alpha_0$ and $\gamma_0$ such that
\begin{equation}
h^y_y(0|\mathfrak{q})=h^t_t(0|\mathfrak{q})\equiv H(\mathfrak{q})
\end{equation}
and use the gauge transformation (\ref{transf}) with $A=-H(\mathfrak{q})/2$ to cancel the transformed metric fluctuations at the boundary. Consequently
\begin{equation}
a_t(0|\mathfrak{q})\to a_t(0|\mathfrak{q})  \qquad a_t^\prime(0|\mathfrak{q})\to a_t^\prime(0|\mathfrak{q})+\frac{1}{2}\mu H(\mathfrak{q})
\end{equation}
and the polarization function $\mathcal{C}_{00}$ reads
\begin{equation}
\mathcal{C}_{00}(0, \mathfrak{q})=\frac{a_t^\prime(0|\mathfrak{q})+\frac{1}{2}\mu H(\mathfrak{q})}{a_t(0|\mathfrak{q})}
\end{equation}
which is evidently a meromorphic function of $\mathfrak{q}$. The theme is proved.

\section{ The WKB approximation at large momentum magnitude }
\label{sec:wkb-approximation-at}

\subsection{The WKB solutions}
\label{sec:wkb-solution}

In the previous section, we obtained the general formula of the electric polarization tensor, $\mathcal{C}_{tt}(\omega, \mathfrak{q})$, in a strongly coupled field theory through the gauge/gravity duality
and linked it to the modified master fields, $\Psi_\pm$ . The expression of $\mathcal{C}_{tt}$, (\ref{eq:36}), in the static limit and its analyticity imply that the  spectrum of singularities of 
electric polarization corresponds to the nontrivial zeros of $a_t$ on the complex momentum-plane.  
While the exact locations of these singularities can only be determined numerically \cite{Blake2015}, the asymptotic distribution of them for a large magnitude of 
momentum can be located analytically through WKB approximation like what we did in our last work for the transverse component of the polarization tensor \cite{Yin2016}. 
As we shall see below, the complex momentum singularities of the electric component share the same Friedel like asymptotic distribution as the transverse component.
Because of the reality of $\mathcal{C}_{tt}(0, \mathfrak{q})$ for real $\mathfrak{q}$, the distribution of the complex singularities is symmetric with respect to the real axis and we need 
only to examine the upper half-plane of the complex $\mathfrak{q}$.

In the master field framework,  all physical quantities involved are taken as  the functions of the modified momentum $k$ through (\ref{eq:37})for convenience. 
We denote the real  and imaginary parts of $k$ as
\begin{align}
  k = w + \mathrm{i} p 
\label{eq:59}
\end{align}
Considering the case where $p \gg |w|$. The ``potential'' of the Schr\"{o}dinger-like equation (\ref{eq:34}) can be approximated as  
\begin{align}
  V_\pm (u |k ) = - \frac{1}{f^2} U_\pm \equiv \frac{Q^2}{f} \left[ - \mathrm{i}k \pm \mathrm{i}u \right]^2 
\label{eq:60}
\end{align}
and the WKB solution reads \footnote{A brief discussion on the WKB approximation and its validity is included in the Appendix C for self-containedness.}
\begin{align}
  \left(\Psi \right)_\pm = \frac{ f^{1/4}} {Q^{1/2}} \, \left( C_1^\pm \; \exp\left\{ Q \int_u^1 \frac{ -k \pm v}{\sqrt{f}} \ \mathrm{d}v \right\} + C_2^\pm \; \exp\left\{ -  Q \int_u^1 \frac{ -k \pm v}{\sqrt{f}} \ \mathrm{d}v \right\} \right)
\label{eq:61}
\end{align}
where $C_1^\pm = C_1^\pm(k)$ and $C_2^\pm = C_2^\pm(k)$ are four constants to be determined.  The difference between (\ref{eq:61}) and the exact solutions are  of the 
order $O(1/p)$, and can be ignored for a large $q$. Matching the WKB solution with the near horizon solution below, the four constants can be reduced to two of them. 

In the limit $u\to 1$, $f(u) \simeq (3 -Q^2)(1 - u)$ as $u \to 1$, the integral in the exponential parts of the WKB solution can be written as
\begin{align}
  \int_u^1 \ \mathrm{d}u \, \frac{Q[-k \pm v]}{\sqrt{f}} \simeq  2\lambda_\pm\sqrt{1- u} , \qquad \text{as} \quad u \rightarrow 0
\end{align}
where 
\begin{align}
  \lambda_\pm \equiv \frac{Q[-k \pm 1]}{\sqrt{3 - Q^2}}  \label{eq:62}
\end{align}
The WKB solution remains approximate in this region as long as $\lambda_\pm\sqrt{1- u} \gg 1$ following from the condition of the approximation: $|V_\pm^\prime| \ll |V_\pm|^{\frac{3}{2}}$ 
and its asymptotic form there is given by
\begin{align}
  \big(\Psi_\pm\big)_{_\text{WKB}} \bigg|_{u \to 1}=  \frac{(3 - Q^2)^{1/4}}{Q^{1/2}} [1-u]^{1/4} \cdot \bigg( C_1^\pm \, e^{ 2 \lambda_\pm \, \sqrt{1-u}} + C_2^\pm \, e^{ - 2 \lambda_\pm\, \sqrt{1-u}} \bigg)  \label{eq:63}
\end{align}
On the other hand,  the Schr\"{o}dinger-like equation (\ref{eq:34}) can be transformed into the modified Bessel equations of the zeroth order, of which the  general solution is given by 
\begin{align}
  \Psi_\pm = \sqrt{1-u} \cdot \bigg[  b_\pm \, I_0(2  \lambda_\pm \sqrt{1-u}) + c_\pm\,  K_0(2  \lambda_\pm \sqrt{1-u}) \bigg]  
\label{eq:64}
\end{align}
with $b_\pm$ and $c_\pm$ constant coefficients. 

For a similar reason in the case of {\it equation (4.10)} in  \cite{Yin2016}, the second term associated to $ K_0(2  \lambda_\pm \sqrt{1-u})$, with a 
logarithmic singularity should be dropped out of (\ref{eq:64}), which means $c_\pm \equiv 0$, in order to prevent the on-shell action  from divergence. Furthermore,  
we are working with the gauge-invariant master field, it's physically legitimate to demand the regularity conditions for the master fields at the horizon. 

In the region $\lambda_\pm\sqrt{1- u} \gg 1$, where both the WKB solutions and the Bessel function solutions approximate, 
we have two forms of the same solution, (\ref{eq:63}) and the asymptotic form of(\ref{eq:64}) at $c_\pm=0$, i.e.
\begin{align}
  \Psi_\pm =  \frac{b_\pm}{2\sqrt{\pi}}\lambda_\pm^{-\frac{1}{2}}(1-u)^{\frac{1}{4}}
  \bigg( e^{ + 2 \lambda_\pm \, \sqrt{1-u}} + \mathrm{i} e^{ -2 \lambda_\pm \, \sqrt{1-u}} \bigg)  
\label{eq:65}
\end{align}
where we have substituted the asymptotic form of $I_0(z)$ for large $|z|$,
\begin{align}
  I_0(z) \simeq \frac{1}{\sqrt{2 \pi z}} \bigg[ \mathrm{i} \, e^{-z}  + e^z   \bigg]  \label{eq:66}
\end{align}
Note that for the limit of $z\to\infty$ along a line parallel to the imaginary axis, both terms inside the bracket of (\ref{eq:66}) have to be retained.
Matching (\ref{eq:64}) and (\ref{eq:65}), we obtain that
\begin{align}
  \mathrm{i}  C_1^\pm = C_2^\pm \equiv \mathrm{i} \, C_\pm  \label{eq:67}
\end{align}
thus the WKB solutions of modified master fields (\ref{eq:63}) become
\begin{equation}
\begin{aligned}
  \big( \Psi_\pm\big)_{_\text{WKB}} = \frac{f^{1/4}}{Q^{1/4}} \, C_\pm \, \bigg[ \exp\left\{  Q \bigg( - k \, \int_u^1 \frac{1}{f(v)}\ \mathrm{d}v \, + (\pm) \int_u^1 \frac{v}{f(v)}\ \mathrm{d}v  \bigg) \right\} \\ + \mathrm{i} \exp\left\{  Q \bigg( k\,  \int_u^1 \frac{1}{f(v)}\ \mathrm{d}v \,   - (\pm) \int_u^1 \frac{v}{f(v)}\ \mathrm{d}v  \bigg) \right\}  \bigg]
  \label{eq:73}
\end{aligned}
\end{equation}

\subsection{ Constraints on the AdS boundary for fluctuations  }
\label{sec:constr-ads-bound}

The WKB solutions obtained in the previous sub-section can be extended all the way to the AdS boundary since $u=0$ ($\zeta=1$) is an ordinary point of the master field 
equation. The background metric near the boundary can be approximated by a pure AdS one. 
In order to extract the electric component of the holographic polarization tensor (\ref{eq:12}), 
we demand that the metric tensor fluctuations vanish on the AdS boundary, i.e. $(h_{\; t}^t, h_{\; x}^x, h_{\; y}^y)\to 0$ as $u\to 0$. 
In this sub-section, we will investigate the asymptotic form of Einstein equations 
(\ref{eq:21})-(\ref{eq:25}) to find a simple approach that leads from the WKB approximation of the master field to the solution for $a_t$ and its 
derivative on the AdS boundary satisfying the above constraints. 

To begin with, the asymptotic form of (\ref{eq:22}) reads
\begin{align}
  \big[ h_{\; t}^t + h_{\; x}^x + h_{\; y}^y\big]^{\prime \prime} - \frac{1}{u}\, \big[ h_{\; t}^t + h_{\; x}^x + h_{\; y}^y\big]^\prime = 0 
\label{eq:51}
\end{align}
which suggests two asymptotic behaviors in the limit $u\to 0$, $\big[ h_{\; t}^t + h_{\; x}^x + h_{\; y}^y \big] \sim 1$ or 
$\big[ h_{\; t}^t + h_{\; x}^x + h_{\; y}^y \big] \sim u^2$. The vanishing limits of the metric fluctuations as $u\to 0$ rules out the first one 
and we have
\begin{equation}
h_{\; x}^x + h_{\; t}^t + h_{\; y}^y = O(u^2)
\end{equation}
Employing this relationship in (\ref{eq:21}), (\ref{eq:23}) and (\ref{eq:24}), we find their identical asymptotic forms, 
\begin{align}
  {h_{\; x}^x}^{\prime \prime} - \frac{2}{u}\, {h_{\; x}^x}^\prime = 0 
\label{eq:68}
\end{align}
and the ones with the index $x$ replaced by $y$ and $t$. Together with the vanishing boundary conditions as $u\to 0$, we end up with
\begin{align}
  h_{\; t}^t , h_{\; x}^x , h_{\; y}^y \sim O(u^3)    \label{eq:70}
\end{align}
It follows from (\ref{eq:70}) and the power series expansion of the eq. (\ref{eq:26}) in $u$ that 
\begin{align}
  \lim_{u \to 0} \, a_{t}'' = Q^2  \mathfrak{q}^2 \lim_{u \to 0 } \, a_t   \label{eq:71}
\end{align}
\begin{align}
  \lim_{u \to 0} \, a_t''' = Q^2 \mathfrak{q}^2 \, \lim_{u \to 0} \, a_t'  \label{eq:72}
\end{align}
The eq. (\ref{eq:38}) and its derivatives link the quantities in (\ref{eq:71}) and (\ref{eq:72}) to the master fields as follows
\begin{align}
  \mathring{a}_t' \equiv \lim_{u \to 0} \, a_t'(u | k) &= \frac{\mu}{2 Q^2}\, \frac{1}{k} \, \left\{ - (Z + k)\, \mathring{\Psi}_- + (Z - k) \mathring{\Psi}_+ \right\}   \label{eq:77}  \\
  \mathring{a}_t'' \equiv  \lim_{u \to 0} \, a_t''(u | k) &= \frac{\mu}{2 Q^2}\, \frac{1}{k}\, \left\{ 2\, \mathring{\Psi}_- - (Z+k) \, \mathring{\Psi}_-' - 2\, \mathring{\Psi}_+ + (Z-k)\,\mathring{\Psi}_+' \right\} \label{eq:78} \\
  \mathring{a}_t''' \equiv \lim_{u \to 0} \, a_t'''(u | k) &= \frac{\mu}{2 Q^2} \, \frac{1}{k} \, \left\{ 4\, \mathring{\Psi}_- - (Z+k)\,\mathring{\Psi}_-'' - 4 \, \mathring{\Psi}_+' + (Z-k) \, \mathring{\Psi}_+'' \right\}  \label{eq:79}
\end{align}
where a ring over a quantity denotes the limit: $u \to 0$. Now we are fully equipped to locate the singularities of $\mathcal{C}_{tt}(0,\mathfrak{q})$ under 
the WKB approximation.

At this point, it is worth comparing the solutions of the master field equations discussed in this section and the solutions of the Einstein-Maxwell equations analyzed 
in the last section. Starting with the gauge potential and metric fluctuations that are analytic at the horizon, we are unable to make all metric fluctuation 
vanishing on the AdS boundary. Employing the residual gauge degrees of freedom, we are able to get rid of all nonzero metric fluctuations on the boundary at the cost of 
introducing non-analytic behavior of the transformed fluctuations. This non-analytic behavior will not show up in the master fields because of their gauge invariance. 

\subsection{ Singularities of electric polarization from WKB approximation }
\label{sec:sing-electr-polar}

Based on the knowledge acquired in previous sub-sections, we are ready to find the spectrum of singularities for the electric polarization  (\ref{eq:36}) at large momentum 
magnitude, which corresponds to the zeros of $a_t$ on the boundary. 

Starting with the WKB solutions \ref{eq:73}), we find its first derivative
\begin{equation}
  \begin{aligned}
    \big( \Psi_\pm\big)_{_\text{WKB}}^\prime  &= f^{1/4} \; \frac{C_\pm}{Q^{1/2}} \,  \bigg[  - Q \frac{ - k \pm u}{\sqrt{f}} \, \exp\left\{  Q \bigg( - k \, \int_u^1 \frac{1}{f(v)}\ \mathrm{d}v \, + (\pm) \int_u^1 \frac{v}{f(v)}\ \mathrm{d}v  \bigg) \right\}  \\
    & \quad  + \mathrm{i} \, Q \frac{-k \pm u}{\sqrt{f}} \, \exp\left\{  Q \bigg( k\,  \int_u^1 \frac{1}{f(v)}\ \mathrm{d}v \,   - (\pm) \int_u^1 \frac{v}{f(v)}\ \mathrm{d}v  \bigg) \right\}  \bigg] + \textit{etc.}
    \label{eq:74}
  \end{aligned}
\end{equation}
where the \textit{etc.} represents all terms that vanish as $u^2$ or faster in the limit $u \to 0$. The second order derivative can be expressed in terms of the modified master fields 
itself through the Schr\"{o}dinger-like equations (\ref{eq:34}).
It follows then that,
\begin{align}
  \big( \mathring{\Psi}_\pm \big)_{_\text{WKB}} \equiv \lim_{u \to 0} \big( \Psi_\pm \big)_{_\text{WKB}} &\simeq C_\pm \; Q^{-1/2}  \;  H_+^{(\pm)}  \label{eq:80} \\
  \big( \mathring{\Psi}_\pm ' \big)_{_\text{WKB}} \equiv \lim_{u \to 0} \big( \Psi_\pm' \big)_{_\text{WKB}} &\simeq C_\pm \;  k Q^{1/2}\, H_-^{(\pm)}  \label{eq:75} \\
  \big( \mathring{\Psi}_\pm '' \big)_{_\text{WKB}} \equiv  \lim_{u \to 0} \big(\Psi_\pm ''\big)_{_\text{WKB}} &\simeq C_\pm \; k^2 Q^{3/2} \, H_+^{(\pm)}   \label{eq:76}
\end{align}
where we retain only the principal term of the "potential" $V_\pm(u | \mathfrak{q})$ (\ref{eq:93}) and (\ref{eq:99}) in accordance with the WKB approximation, 
\begin{align}
  \big( \lim_{u \to 0} V_\pm + Q^2 \mathfrak{q}^2\big)_{_\text{WKB}} = 0    \label{eq:53}
\end{align}
and introduce the notations:
\begin{equation}
  \begin{aligned}
    H_-^{(\pm)} &\equiv \mathrm{e}^{  Q ( - k \, L_1\pm L_2  ) } - \mathrm{i} \mathrm{e}^{  Q ( k\,  L_1 - (\pm) L_2 ) }  \\
    H_+^{(\pm)} &\equiv\mathrm{e}^{  Q ( - k \, L_1\pm L_2  ) } + \mathrm{i} \mathrm{e}^{  Q ( k\,  L_1  - (\pm) L_2 ) } 
\label{eq:87}
  \end{aligned}
\end{equation}
for brevity with $L_1$ and $L_2$ denoting the two elliptic integrals 
\begin{equation}
\begin{aligned}
  L_1  &\equiv \int_0^1 \frac{1}{f(v)}\ \mathrm{d}v  \\
  L_2  &\equiv  \int_0^1 \frac{v}{f(v)}\ \mathrm{d}v
  \label{eq:40}
\end{aligned}
\end{equation}

According to condition (\ref{eq:71}) together with (\ref{eq:78})(\ref{eq:80}) and (\ref{eq:75})  , we obtain $\lim\limits_{u \to 0} a_t$ in WKB approximation: 
\begin{align}
  \big( \mathring{a}_t \big)_{_\text{WKB}} \simeq \frac{1}{Q^2 k^2} \big( \mathring{a}_t '' \big)_{_\text{WKB}} = - \frac{\mu}{2 Q^2} \, \frac{Q^{-3/2}}{k} \; \bigg[  H_-^{(-)} \; C_- + H_-^{(+)} \; C_+  \bigg]   \label{eq:81}
\end{align}
As its zeros give rise to the poles of $\big( \mathcal{C}_{tt}\big)_{_\text{WKB}}$, we find one of the equations for their locations: 
\begin{align}
\big(\mathring{a}_t  \big)_{_\text{WKB}} \sim H_-^{(-)} \; C_- + H_-^{(+)} \; C_+ = 0   \label{eq:83}
\end{align}
Substituting (\ref{eq:77}) and (\ref{eq:79}) into (\ref{eq:72}) with $\mathring{\Psi}_\pm$, $\mathring{\Psi}_\pm'$ and $\mathring{\Psi}_\pm ''$ given by their WKB approximations 
(\ref{eq:80}), (\ref{eq:75}) and (\ref{eq:76}), we find another equation for the poles,
\begin{align}
  H_-^{(-)} \; C_- - H_-^{(+)} \; C_+ = 0   \label{eq:85}
\end{align}
Equations (\ref{eq:83}) and (\ref{eq:85}), form a system of linear homogeneous equations in two unknowns coefficients ,$C_-$ and $C_+$, the existence of nontrivial solutions for them implies that
\begin{align}
  0 = 
\begin{vmatrix}
H_-^{(-)} & H_-^{(+)} \\
H_-^{(-)}  & - H_-^{(+)}
\end{vmatrix}  
= -2 \, H_-^{(-)} \, H_-^{(+)}
\label{eq:86}
\end{align}
Substituting the explicit expressions of $H_-^{(-)}$ and $H_-^{(+)}$ in (\ref{eq:87}) and (\ref{eq:40}), we obtain 
\begin{align}
  \mathrm{e}^{-2 Q \, L_1 k} - \mathrm{i} \mathrm{e}^{-2 Q L_2} - \mathrm{i} \mathrm{e}^{2 Q L_2} - \mathrm{e}^{2 Q L_1 k} = 0 
  \label{eq:88}
\end{align}
the solution of (\ref{eq:88}) in the complex-plane is given by
\begin{equation}
    \left( \mathrm{e}^{-2 Q L_1 w}  - \mathrm{e}^{2 Q L_1 w} \right) \, \cos\big[  2 Q L_1 p \big] = 0   \label{eq:89}  
\end{equation}
and
\begin{equation}
    -\frac{\mathrm{e}^{- 2 Q L_2} + \mathrm{e}^{2 Q L_2}}{\mathrm{e}^{-2 Q L_1 w} + \mathrm{e}^{2 Q L_1 w}}  =   \sin \big[ 2 Q L_1 p \big]  \label{eq:90}
\end{equation}
Finally, from (\ref{eq:89}) and (\ref{eq:90}), we obtain the positions of the singularities of the modified momentum $k = w + \mathrm{i}p$ for $p \gg |w|$, i.e.,
\begin{align}
  w &= \pm \frac{L_2}{L_1}  \label{eq:91} \\
  p &= \frac{\pi}{Q L_1} \big[ n - \frac{1}{4} \big]  \label{eq:92}
\end{align}
It follows from (\ref{eq:37}) that $\mathfrak{q}\simeq k$ for $k \gg |w|$ and the asymptotic locations of the poles of $\mathcal{C}(0,\mathfrak{q})$ on the upper complex $\mathfrak{q}$ 
plane are thereby 
\begin{equation}
\mathfrak{q}\simeq \pm w + i \frac{\pi}{Q L_1} \big[ n - \frac{1}{4} \big]
\label{final}
\end{equation}
The asymptotic locations of the poles on the lower complex $\mathfrak{q}$ plane are obtained from (\ref{final}) by a reflection with respect to the real axis.

\section{ Concluding Remarks }

Let us recapitulate what we did in this work. We started with an one-loop calculation of the static electric component of the polarization tensor of a spinor 
QED3 and a scalar QED3 with chemical potential $\mu$ and explored its analyticity on the complex momentum plane. We found an infinite number of branch points along two straight-lines parallel 
to the imaginary axis with real part equal to $\pm\mu$. The nature of these singularities and their distribution serve a benchmark of the holographic polarization tensor discussed in the 
subsequent sections and the ones close to the real axis are responsible to the Friedel oscillations observed in some materials. Then we explored the analyticity of the static electric
component of the holographic polarization tensor and proved that it is a meromorphic function of the complex momentum. Employing the WKB approximation, we are able to 
locate analytically the asymptotic distributions of the poles along two lines parallel to the imaginary axis of the momentum plane, i.e.
\begin{equation}
\frac{q}{\mu}=\pm\frac{L_2}{L_1}\pm i\frac{\pi}{QL_1}|n-\frac{1}{4}|   \label{eq:69}
\end{equation}
for $n \gg 1$.

It would be interesting to compare the asymptotic locations of the poles (\ref{eq:91})and (\ref{eq:92}) with the those extracted from the numerical solutions of the Einstein-
Maxwell equations reported in \cite{Blake2015}. Comparing our Einstein-Maxwell equations with those in the Appendix A of \cite{Blake2015}, 
we noted the following relationships between our notations and theirs \footnote{Throughout this paper, we have scaled the $U(1)$ gauge potential in eq.(\ref{eq:13}) such that
$\frac{K_4}{G_4}=L^2=1$. An arbitrary ratio $\frac{K_4}{G_4}\equiv\eta^2$ amounts to the transformations $a_t\to\eta a_t$ and $\mu\to\eta\mu$ in the Einstein-Maxwell 
equations (\ref{eq:21})-(\ref{eq:29}). While eq. (\ref{eq:69}) is $\frac{q}{\mu}=\pm\frac{L_2}{L_1} \eta \pm \mathrm{i}\frac{\pi}{QL_1} \eta |n-\frac{1}{4}|$ . The notation in \cite{Blake2015} corresponds to $\eta=\frac{1}{2}$. }

\begin{table}[ht]
  \centering
  \begin{tabular}{c | c}
    \hline\hline
    Our notation & The notation of \cite{Blake2015}\\ 
    \hline
    $a_t$ & $\delta A_t/2$\\
    $h^t_t$ & $\delta g^t_t$\\
    $h^x_x$ & $\delta g^x_x$\\
    $h^y_y$ & $\delta g^y_y$\\
    $\mu$ & $\mu_0/2$\\
    $q$ & $k$\\
    $u$ & $z/z_+$\\
    \hline  
  \end{tabular}  
  \caption{The Translation of Notations}
\label{tab:fig}
\end{table}

It follows from Table \ref{tab:fig}  that
\begin{equation}
  \frac{k}{\mu_0}=\frac{q}{2\mu}
\end{equation}
According to eqs. (\ref{eq:91})and (\ref{eq:92}), we find the poles at
\begin{equation} 
  \frac{k}{\mu_0} = \pm 0.309558 + 2.12399 \,  \left(n - \frac{1}{4} \right) \mathrm{i} \label{eq:109}
\end{equation} 
 for $\frac{T}{\mu_0} = 0.21$ and 
\begin{equation} 
  \frac{k}{\mu_0} =\pm 0.399722 + 0.288434 \, \left(n - \frac{1}{4} \right) \mathrm{i} \label{eq:110}
\end{equation}
for $\frac{T}{\mu_0} =0.0006$, here $n \in \mathbb{Z}$. Moreover,  the  equations (\ref{eq:109}) and (\ref{eq:110})  indicate that  the separation between any two nearest 
poles is $2.12399$ and $0.288434$  for the two cases. 
These results are very close to the numerically determined locations in Figure 7 of \cite{Blake2015}. We can see that our WKB-type  formalism holds in a large imaginary 
part of complex momentum in principle, which means integer parameters $n$ in  (\ref{eq:109}) and (\ref{eq:110}) should be large enough as to the coefficient $\frac{1}{4}$ 
can be neglected asymptotically.  However, comparing with these numerical simulation, even in a regime not too far away from the real axis, our WKB approximations
works reasonably well. 

Our study on the analyticity of the  electric component of the static polarization tensor is consistent with the previous knowledge on the zero chemical potential case \cite{Hou2010}.  In the absence of the chemical potential, all of the Friedel-like poles migrate to the imaginary axis, and it was also proved that the static polarization tensor is a meromorphic function.  The existence of these poles 
can be inferred from the Matsubara formulation of the boundary field theory, whose Lagrangian density is $O(3)$ invariant at $\mu=0$. Mathematically, it is optional to 
interpret any dimension as Euclidean time with the associated momentum as the imaginary energy. Therefore the poles on the imaginary $q$-axis corresponds to the excitation 
of another boundary field theory with the original compact Euclidean time and one of the spatial dimension interpreted as a compact space dimension and Euclidean time. 

The WKB approximation can be readily applied when the momentum $q$ is real and large. Up to the leading order, the WKB solution (\ref{eq:61}) takes the form
\begin{equation}
  \left(\Psi \right)_\pm \simeq  \mathrm{const.}\exp\left(|q|z_+ \int_u^1 \frac{1}{\sqrt{f}} \ \mathrm{d}v \right) 
\end{equation}
in this case, where we have dropped the exponentially small term, identified $k$ with $q/\mu$ and substituted $Q=\mu z_+$. It follows from (\ref{eq:36}) and (\ref{eq:71}) 
together with the leading order of (\ref{eq:77}) and (\ref{eq:79}) that
\begin{equation}
\mathcal{C}_{00}(0,q)= K_4 q
\end{equation}
The UV behavior is not impacted by the nonzero temperature and chemical potential as expected. Notice that because of the low dimensionality and gauge invariance, 
the polarization tensor is free from UV divergence.

The presence of the Friedel-like singularities appears a common property of a quantum field theory (strongly or weakly coupled) with a nonzero chemical potential.  
It is not yet a sufficient evidence of the fermionic degrees of freedom in the boundary field theory of the gravity dual. More investigations are warranted to probe
the fermionic degrees of freedom in a holography implied field theory, for example by adding a spinor field in a bulk theory \cite{Lee2009, Basu2010b, liu2011, mihailo2011,faulkner2011, gubser2012}.


\appendix
\section{The master field equations in even parity}
\label{sec:mast-field-equat}

The function $\alpha_\pm$ in the definition of the master fields (\ref{eq:31}) is given by 
\begin{align}
  \alpha_\pm &\equiv  \frac{Q^2}{2}\big[ Z \pm \sqrt{Z^2 + \mathfrak{q}} \big] - Q^2 \, u   \label{eq:95} \\
\end{align}
with $Z$ defined in (\ref{eq:39}).

The explicit expression  of $U_\pm$ in the ``potential'' of the Schr\"{o}dinger-like equation (\ref{eq:93}) reads 
\begin{equation}
  \begin{aligned}
    U_\pm &\equiv  \pm f \; \Bigg\{ \frac{Z - 2u \pm k}{2k} \cdot V + \frac{Q^2 k^4 - [ \frac{f'}{u} + 2 Z] k^2 + Q^2 Z^4 + \frac{f'}{u}\, Z^2 + 8 f}{2k ( Z - 2 u \pm k)} \\
    & \hspace{4cm} - \frac{2u (Q^2 k^2 - Q^2 Z + \frac{f'}{u})}{k} + \frac{4 f [ Q^2 (8u-3) -3 ]}{k \, ( Q^2 k^2 - Q^2 Z^2 - \frac{f'}{u}) } \Bigg\}
    \label{eq:96}
  \end{aligned}
\end{equation}
where
\begin{equation}
  \begin{aligned}
    V &\equiv \frac{1}{\big[ Q^2 u  \, k ^2 - Q^2 Z^2 \, u - f'\big]^2 } \bigg\{ Q^6 u^2 \cdot \mathfrak{q}^6 + 3 \, Q^4 (1+Q^2) u^3 \cdot \mathfrak{q}^4  \\
    & \quad  + \big[  8 \,  Q^2 u^2 - 153 \, Q^2 (1 +Q^2)^2 u^4 + 400 \,  Q^4 (1 + Q^2) u^5 + 264 \, Q^6 u ^6 \big] \cdot \mathfrak{q}^2  \\
    & \quad  + u^2 \big[ - 738 \,  (1 + Q^2)^2 + 1944 \,  Q^2 (1 + Q^2) u - 1248 \, Q^4 u^2 + 1197 ( 1 + Q^2)^3 u ^3  \\
    & \quad  - 4482 \, Q^2 ( 1 +Q^2 )^2 u ^4 + 5544 \,  Q^4 ( 1 + Q^2) u ^5 - 2272 \, Q^6 u ^6 \big] \bigg\}
    \label{eq:97}
  \end{aligned}
\end{equation}
From the explicit expressions of $U_\pm$, (\ref{eq:96}) and (\ref{eq:97}), it's easy to verify that the AdS boundary $u=0$ is an ordinary point of the 
master field equations (\ref{eq:32}) because none of its coefficients diverges there. 

For a large momentum magnitude, the dimensionless momentum is approximately equal to modified momentum: i. e. $\mathfrak{q} \sim k$, and
\begin{equation}
  \begin{aligned}
    V &= \frac{Q^4 \mathfrak{q}^4 + 4 \, u^2 Q ^4 \mathfrak{q}^2}{ Q^2 \mathfrak{q}^2 + 3\, u ( 1 - u Q^2) } + O\left(\frac{1}{\mathfrak{q}^2} \right)    \\
    &= Q^2 \, k^2 + Q^2 (4 \, u^2 - Z^2) + O\left(\frac{1}{\mathfrak{q}^2} \right) \\
    \label{eq:98}
  \end{aligned}
\end{equation}
Substituting (\ref{eq:98}) into (\ref{eq:97}), after some algebra, we obtain 
\begin{equation}
  \begin{aligned}
    U_\pm = f\, \bigg\{ Q^2 [ k \mp u ]^2 + u^2 Q^2 - \frac{1}{2} Q^2 Z + O\left( \frac{1}{k} \right) \bigg\}
    \label{eq:99}
  \end{aligned}
\end{equation}
Only the leading and sub-leading terms of (\ref{eq:99}) in large $k$ are required to carry out the WKB approximation for our purpose
and we may write
\begin{equation}
  U_\pm \simeq f \, Q^2 [ k \mp u]^2 = - f \, Q^2 [ - \mathrm{i} \,  k \pm \mathrm{i} \, u ]^2
\end{equation}
which turns into (\ref{eq:60}), and  $O\big(\frac{1}{\mathfrak{q}} \big) = O\left( \frac{1}{k} \right)$ holds in the WKB approximation.


\section{The proof of the convergence of the power series solutions for $a_t, h_{\, t}^t , h_{\, y}^y$}
\label{app:convergence}

The explicit recursion formulas for the coefficients of the power series solution of (\ref{eq:49}), (\ref{eq:50}) and (\ref{eq:52}), 
following from the eqs. (\ref{eq:46}), (\ref{eq:47}) and (\ref{eq:48}) read
\begin{equation}
  \begin{aligned}
    \gamma_{n+1} &= - \frac{ 1 }{ (n+1) (n + c^{(0)}_5) } \sum\limits_{k=0}^n \bigg[ (k+1) \, c_3^{(n-k)} \, \alpha_k  + \big( k \, c_4^{(n-k)} + d_6^{(n-k)} \big) \, \beta_k  \\
    & \hspace{7cm}  + \big( k \, c_5^{(n-k+1)} + d_7^{(n-k)} \big) \, \gamma_ k \bigg]
  \label{eq:B1}
  \end{aligned}
\end{equation}
\begin{equation}
  \begin{aligned}
    \beta_{n+1} &= \frac{ 1 }{ ( n + c_5^{(0)})( n + 1 d_2^{(0)}) } \sum\limits_{k=0}^n \bigg[ (k+1)\, c_3^{(n-k)} \, \alpha_k + \big( k \, c_4^{(n-k)} + d_6^{(n-k)} \big) \, \beta_k \\
    &\qquad + \big( k \, c_5^{(n-k+1)} + d_7^{(n-k)} \big) \, \gamma_k \bigg] - \frac{ 1 }{ n + 1 + d_2^{(0)} } \sum\limits_{k=0}^\infty \bigg[ d_1^{(n-k)} \, a_k + d_2^{(n-k+1)} \, \beta_k \bigg]
  \label{eq:B2}
  \end{aligned}
\end{equation}
\begin{equation}
  \begin{aligned}
    \alpha_{n+1} &= \sum\limits_{k=0}^n \Bigg\{ \bigg[ \frac{ c_2^{(0)} \cdot d_1^{(n-k)} }{ ( n+1+d_2^{0})(n+2) } - \frac{ c_2^{(0)} \cdot (k+1) \, c_3^{(n-k)} }{ (n + c_5^{(0)})(n+1+d_2^{(0)})(n+2) }  \\
& \hspace{7cm}   - \frac{ (k+1)\, c_1^{(n-k)} + d_3^{(n-k)} }{ ( n + 1 )( n + 3 ) } \bigg] \cdot  \alpha_k \\
    & + \bigg[ \frac{ c_2^{(0)} }{ ( n + 1 + d_2^{(0)})( n + 2) } \bigg( d_2^{(n-k+1)} - \frac{ k \, c_4^{(n-k)} + c_6^{(n-k)} }{  n + c_5^{(0)} }  \bigg) - \frac{ k \, c_2^{(n-k+1)} + d_4^{(n-k)}}{ (n+1)(n+2) } \bigg] \cdot \beta_k  \\
    &+ \bigg[ - \frac{ c_2^{0} \cdot \big( k \, c_5^{(n-k+1)} + d_2^{(n-k)} \big) }{ ( n + c_5^{(0)})( n+ 1 + d_2^{(0)}) } - \frac{ d_5^{(n-k)} }{ ( n+ 1)( n + 2) }  \bigg] \cdot \gamma_k \Bigg\}
  \label{eq:B3}
  \end{aligned}
\end{equation}
where $c_i^{(m)}$'s and $d_i^{(m)}$'s are defined by the Taylor expansions of the coefficient functions $C_i$'s and $D_i$'s in (\ref{eq:45}). We have
\begin{align}
  c_1^{(0)} &=  \frac{ 4 Q^2 }{ 3 - Q^2 }  &\quad 
  c_2^{(0)} &= - \frac{ 1 }{ 2 }  \\
  c_3^{(0)} &= \frac{ 4 }{ 3 - Q^2 }  &\quad 
  c_4^{(0)} &= 1 \\
  c_5^{(0)} &= 1 &  & \\
  d_1^{(0)} &= - \frac{ 4 Q^2 }{ 3 - Q^2 }  &\quad d_2^{(0)} &=  \frac{ 1 }{ 2 }  \\
  d_3^{(0)} &= - \frac{ Q^2 }{ 3 - Q^2 } \mathfrak{q}^2 &\quad 
  d_4^{(0)} &= - \frac{  Q^2 \big(  \mathfrak{q}^2 + 2 \big) }{ 3 - Q^2 }  \\
  d_5^{(0)} &= - \frac{  Q^2 \, \mathfrak{q}^2 }{ 3 - Q^2 }  &\quad d_6^{(0)} &= - \frac{ 2 Q^2 }{ 3 - Q^2 }  \\
  d_7^{(0)} &= - \frac{ Q^2 }{ 3 - Q^2 }  & &  
\end{align}
The coefficients $\alpha_n$, $\beta_n$ and $\gamma_n$ can be bounded following the induction method employed in \cite{Whittaker1952} for the power series solution
of an ordinary differential equation around a regular point.
The analyticity of $C_i$ and $D_i$ at the horizon $\zeta = 0$ implies that there exists a circular domain: $|\zeta| < r$, where 
all functions of $C_i, D_i$ are bound and satisfy the Cauchy inequality
\begin{align}
  | c_i^{(k)} | < \frac{M}{8r^k}  &\qquad     | d_i^{(k)} | < \frac{M}{8r^k}, \qquad      | c_i^{(k)} + d_j^{(k)}   | < \frac{M}{8r^k} \\ 
  | \alpha_0 | \le\frac{M}{r}  &\qquad         | \beta_0 | = 0  \le  \frac{M}{r}   \qquad       | \gamma_0 |  \le \frac{M}{r} 
\end{align}
for a finite constant $M$. We choose $r<1$ and $M>1$ and start with the inequalities
\begin{align}
  | \alpha_0 | \le\frac{M}{r}  \qquad         
  | \beta_0 | = 0  \le  \frac{M}{r}   \qquad       | \gamma_0 |  \le \frac{M}{r} 
\end{align}
for given $\alpha_0$ and $\gamma_0$.
Assuming that all $\alpha_k$, $\beta_k$ and $\gamma_k$ with $k \le n$ are known and satisfy the inequalities
\begin{equation}
  | \alpha_k | \le \left( \frac{M}{r} \right)^k \qquad   | \beta_k | \le \left( \frac{M}{r} \right)^k \qquad  | \gamma_k | \le \left( \frac{M}{r} \right)^k
\label{eq:B14}
\end{equation}
we shall prove that
\begin{equation}
  | \alpha_{n+1}| \le \left( \frac{M}{r} \right)^{n+1} \qquad | \beta_{n+1} | \le \left( \frac{M}{r} \right)^{n+1} \qquad | \gamma_{n+1}  | \le \left( \frac{M}{r} \right)^{n+1}
\end{equation}

Following (\ref{eq:B1}), we obtain that
\begin{equation}
  \begin{aligned}
    | \gamma_{n+1} |  & \le  \frac{1}{(n+1)(n+c_5^{(0)})}\bigg\{ \sum_{k=0}^n(k+1)   |c_3|^{n-k}| \alpha_k | + \sum_{k=0}^n\left(k |c_4^{n-k}|+|d_6^{(n-k)}|\right)|\beta_k|   \\ 
    &  \hspace{4cm} +\sum_{k=0}^n\left(k |c_5^{n-k+1}|+|d_7^{(n-k)}|\right)|\gamma_k| \bigg\}  \\
    &\le \frac{M}{8(n+1)(n+c_5^{(0)}) \, r^{n+1}}\Big[r \sum_{k=0}^n(k+1)M^k+r\sum_{k=0}^n(k+1)M^k+\sum_{k=0}^n\, (k+1) M^k \Big]\\
    & \le  \frac{3 (n+2)}{ 16(n+c_5^{(0)} ) }\left(\frac{M}{r}\right)^{n+1} \\
    &\le \frac{3}{8}\left(\frac{M}{r}\right)^{n+1} \le \left(\frac{M}{r}\right)^{n+1}
    \label{eq_1}
  \end{aligned}
\end{equation}

where we have used $c_5^{(0)}=1$ in the last step. Then (\ref{eq:47})  and the  $d_2^{(0)}=\frac{ 1 }{ 2 }$ yield
\begin{equation}
  \begin{aligned}
    | \beta_{n+1} | & \le   | \gamma_{n+1} |+\frac{1}{n+1} \sum_{k=0}^n \left(  | d_1^{(n-k)} | \, | \alpha_k | + | d_2^{(n-k+1)} | \, | \beta_k | \right) \\
    & \le  \frac{3}{8} \left( \frac{M}{r} \right)^{n+1} + \frac{1}{8} (r+1)\left( \frac{M}{r} \right)^{n+1} \\
    &\le \frac{5}{8} \left( \frac{M}{r} \right)^{n+1} \le \left(\frac{M}{r}\right)^{n+1}
    \label{eq_2}
  \end{aligned}
\end{equation}
Finally, it follows from (\ref{eq:48})  that
\begin{equation}
  \begin{aligned}
    |\alpha_{n+1}| & \le  \frac{ | c_2^{(0)}  | }{n+2} |\beta_{n+1}|+\frac{1}{(n+1)(n+2)}\sum_{k=0}^n \bigg\{ \Big[k |c_1^{(n-k)}|+|c_1^{(n-k)}+|d_3^{(n-k)}||\Big]|\alpha_k| \\
    & \hspace{3cm}  + \Big[k|c_2^{(n-k+1)}|+|d_4^{(n-k)}|\Big]|\beta_k|+|d_5^{(n-k)}||\gamma_k| \bigg\} \\
    &\le \frac{5}{8}\left(\frac{M}{r}\right)^{n+1}+\frac{(r+2)}{8}\left(\frac{M}{r}\right)^{n+1} \le \left(\frac{M}{r}\right)^{n+1}
    \label{eq_3}
  \end{aligned}
\end{equation}
The theme is then proved and the inequalities (\ref{eq:B14}) hold for any $k$. 
It follows from eq.(\ref{eq:55}) that
\begin{eqnarray}
  |a_t| &\le& \mu |\zeta|\sum_{n=0}^\infty|\alpha_n||\zeta|^n\le\mu |\zeta|\sum_{n=0}^\infty\left(\frac{M|\zeta|}{r}\right)^n
  =\frac{\mu|\zeta|}{1-\frac{M|\zeta|}{r}}\nonumber \\
  |h^t_t| &\le& \sum_{n=0}^\infty|\beta_n||\zeta|^n\le\sum_{n=0}^\infty\left(\frac{M|\zeta|}{r}\right)^n
  =\frac{1}{1-\frac{M|\zeta|}{r}}\nonumber \\
  |h^y_y| &\le& \sum_{n=0}^\infty|\gamma_n||\zeta|^n\le\sum_{n=0}^\infty\left(\frac{M|\zeta|}{r}\right)^n
  =\frac{1}{1-\frac{M|\zeta|}{r}}
\end{eqnarray}
The series solution (\ref{eq:49})-(\ref{eq:52}) is therefore convergent for $|\zeta|<\frac{r}{M}$ and is analytic there. Since the coefficients of the recursion formulas 
(\ref{eq:B1}), (\ref{eq:B2}) and (\ref{eq:B3}) are all polynomials of $q$, eqs.(\ref{eq:49})-(\ref{eq:52}) also define three analytic functions of $q$ for $\zeta$ within the convergence circle.

Next we prove the absence of diverging solutions at AdS boundary, $u =0$ ($\zeta=1$). To see this, we write down the asymptotic form of the Einstein-Maxwell equations (\ref{eq:44})  near the 
boundary by retaining only the leading term of the power series of the coefficient functions $C_i$ and $D_i$ in $u$ .
\begin{align}
  0 &= {h^t_t}^\prime + {h^y_y}^\prime +p_1u^2 a_t + p_2 u^2 h^t_t  \label{eq1}\\
  0 &= a_t^{\prime\prime}+p_3 u^3 a_t^\prime + p_4 {h^t_t}^\prime + p_5 a_t + p_6 u h^t_t + p_7 u h^y_y  \label{eq2}  \\
  0 &= {h^y_y}^{\prime\prime} - \frac{2}{u} {h^y_y}^\prime+p_8 u^2 a_t^\prime + p_9 u^2  {h^t_t}^\prime + p_{10} h^t_t + p_{11} h^y_y    \label{eq3} \\
\end{align}
with $p_1,...,p_{11}$ numerical constants. The terms dropped in (\ref{eq3}) will not contribute to the analysis below. Substituting the power series solution
\begin{equation}
  \begin{aligned}
    a_t &= a_0+a_1u+a_2u^2+a_3u^3+... \\
    h^t_t &= b_0+b_1u+b_2u^2+b_3u^3+... \\
    h^y_y &= c_0+c_1u+c_2u^2+c_3u^3+...
    \label{eq4}
  \end{aligned}
\end{equation}
into (\ref{eq1}), (\ref{eq2}) and (\ref{eq3}) , we obtain

\begin{equation}
  \begin{aligned}
    0 &= b_1+c_1 \\
    0 &= b_2+c_2  \\
    0 &= 3(b_3+c_3)+p_1a_0+p_2b_0
  \end{aligned}
\end{equation}
from (\ref{eq1}); and
\begin{equation}
  \begin{aligned}
    0 &= 2a_2+p_4b_1+p_5a_0  \\
    0 &= 6a_3+2p_4b_2+p_5a_1+p_6b_0+p_7c_0
  \end{aligned}
\end{equation}
from (\ref{eq2}) , and
\begin{equation}
  \begin{aligned}
    0 &= c_1  \\
    0 &= -2c_2+p_{10}b_0+p_{11}c_0  \\
    0 &= p_{10}b_1+p_{11}c_1
  \end{aligned}
\end{equation}
from (\ref{eq3}). Consequently, we find that
\begin{equation}
  \begin{aligned}
    b_1 &= c_1 = 0 \\
    a_2 &= -\frac{1}{2} p_5 a_0 \\
    c_2 &= \frac{1}{2}(p_{10}b_0+p_{11}c_0)  \\
    b_2 &= -c_2=-\frac{1}{2}(p_{10}b_0+p_{11}c_0) \\
    a_3 &= \frac{1}{6}\left[(p_4p_{10}-p_6)b_0+(p_4p_{11}-p_7)c_0+p_5a_1\right]   \\
    c_3 &= -b_3 -\frac{1}{3}(p_1 a_0 + p_2 b_0)
  \end{aligned}
\end{equation}
Iteratively, the five free parameters on RHS, $a_0$, $a_1$, $b_0$, $c_0$ and $b_3$, covers all solutions near the boundary. None of them and their derivatives 
diverge at $u=0$. It follows that the series solution (\ref{eq:49})-(\ref{eq:52}) or the analytic continuation in (\ref{eq:55}) and its derivative with respect to $u$, 
remains convergent on the AdS boundary, and therefore are analytic functions of $q$ at $u=0$ following the Weierstrass theorem.

\section{Validity Domain for the WKB approximation}
\label{sec:validity-domain-wkb}
  Consider a Schr\"{o}dinger-like equation
  \begin{equation}
    \frac{\mathrm{d}^2\Psi}{\mathrm{d} u^2} - V \Psi = 0
    \label{Schroedinger}
  \end{equation}
  where the complex potential is parametrized as 
  \begin{equation}
    V(u) = (\lambda a + b)^2 + c
  \end{equation}
  where $\lambda \gg 1$, ($a$, $b$) are functions of $u$, independent of $\lambda$, and $c$ is a function of $u$ and $\lambda$ but is of the $O(1)$ in $\lambda$. 
To the order of approximation made in this work, the function $c$ may be dropped.  
 
On writing
  \begin{equation}
    \Psi(u) = \mathrm{e}^{S(u)}
  \end{equation}
  the differential equation satisfied by $S(u)$ reads
  \begin{equation}
    \frac{\mathrm{d}^2S}{ \mathrm{d} u^2}+\left(\frac{ \mathrm{d}S}{ \mathrm{d} u}\right)^2-(\lambda a+b)^2 - c = 0
    \label{HJ}
  \end{equation}
  Let 
  \begin{equation}
    S = \lambda S_0 + S_1 + O \left(\frac{1}{\lambda} \right) + \cdots
  \end{equation}
  the eq.(\ref{HJ}) takes the form 
  \begin{equation}
    \lambda^2 \left[ \left(\frac{\mathrm{d}S_0}{\mathrm{d}u} \right)^2  -  a^2 \right]+\lambda\left(\frac{\mathrm{d}^2S_0}{\mathrm{d} u^2}-2i\frac{\mathrm{d}S_0}{\mathrm{d}u}\frac{\mathrm{d}S_1}{\mathrm{d}u}+2 \mathrm{i} ab \right) = O(1)
  \end{equation}
  upto $O(1)$ in $\lambda$, which boils down to the following equations:
  \begin{eqnarray}
    \left( \frac{ \mathrm{d}S_0}{ \mathrm{d}u} \right)^2-a^2  &=& 0 \notag \\
    \mathrm{i} \frac{\mathrm{d}^2S_0}{ \mathrm{d}u^2}+2\frac{\mathrm{d}S_0}{\mathrm{d}u} \frac{\mathrm{d}S_1}{ \mathrm{d}u} - 2 ab &=& 0
  \end{eqnarray}
It follows that
  \begin{equation}
    \begin{aligned}
    \frac{\mathrm{d} S_0}{\mathrm{d}u} &= \pm a  \\
    \frac{\mathrm{d}S_1}{ \mathrm{d} u} &= -\frac{1}{2}\frac{\mathrm{d}}{\mathrm{d}u}\ln a+b.
  \end{aligned}
\end{equation}
  The WKB solution is then
  \begin{equation}
    \Psi(u) \simeq = \frac{1}{ \sqrt{a(u)} } \mathrm{e}^{\pm\int_0^u \ \mathrm{d}u'[\lambda a(u')+b(u')] }
  \end{equation}
  and its difference from the exact solution is $O(1/\lambda)$. 

In this work, $\lambda$ corresponds to the imaginary part of the momentum. 
  It is well-known that the WKB approximation fails near the zero of the potential (turning point), which does not happen here. The approximation also breaks down near the singularity of 
  the potential where the $O(\lambda)$ terms in (\ref{HJ}) diverges faster than $O(\lambda^2)$ terms. Therefore the validity of the WKB requires
  that $\lambda a \gg \frac{\ \mathrm{d}a}{\mathrm{d} u}$, which, in our case with $a=\frac{1}{\sqrt{f}}$, excludes the region arbitrarily close to the horizon.

\subsection*{Acknowledgments}
\label{sec:acknowledgments}

D.F.H. and H.C.R. are supported by by the Ministry of Science and Technology of China (MSTC) under the ''973''  Project No. 2015CB856904. L.Y. and T.K.L. acknowledge the support by MOST 104-2112-M-001-005. And L.Y. also supported by QLPL2015P01 under No. 201508; D.F.H. and H.C.R. by NSFC under Grants No. 11375070, No. 11521064 and No. 11135011.

\bibliographystyle{JHEP}

\providecommand{\href}[2]{#2}\begingroup\raggedright
\endgroup

\end{document}